\documentclass[11pt,letterpaper]{article}
\usepackage[margin=1in]{geometry}

\RequirePackage{amsthm,amsmath,amsfonts,amssymb}
\RequirePackage[authoryear,round]{natbib}
\RequirePackage{graphicx}
\RequirePackage{pdfpages}
\RequirePackage{booktabs}
\RequirePackage{tabularx}
\RequirePackage{multirow}
\RequirePackage{subcaption}
\RequirePackage{url}

\begin{document}

\title{Hierarchical Bayesian Joint Modeling of Athlete Performance Trajectories and Latent Peak Ages}
\author{K.\ Ken Peng$^{1,2}$ \qquad Tianyu Guan$^3$ \qquad Xiong Zhao$^{4,5}$\\[0.7em]
\parbox{0.92\textwidth}{\centering\small
$^1$Department of Statistics and Actuarial Science, Simon Fraser University, Burnaby, British Columbia, Canada\\
$^2$Department of Civil Engineering, University of Ottawa, Ottawa, Ontario, Canada\\
$^3$Department of Mathematics and Statistics, York University, Toronto, Ontario, Canada\\
$^4$AccMov Health Inc., Ottawa, Ontario, Canada\\
$^5$School of Kinesiology and Health Studies, Queen's University, Kingston, Ontario, Canada}}
\date{}
\maketitle

\begin{abstract}
Determining when athletes reach peak performance is a central question in sports science, with implications for talent identification, training design, and cross-discipline comparisons. Statistical inference is challenging because peak age is not directly observed: competition records are noisy and selective measurements of underlying performance capacity and are often limited to incomplete observation windows. We develop a hierarchical Bayesian joint modeling framework that links a longitudinal model for athlete-specific performance trajectories to a regression submodel for latent peak age. The longitudinal component models observed competition performances as noisy measurements of athlete-specific performance trajectories over age. The peak-age component then treats the maximizer of each trajectory as a latent response in a parametric accelerated failure time (AFT) model, allowing athlete-level covariates to explain earlier or later peak age. The hierarchical structure borrows information across athletes, while joint posterior inference propagates uncertainty from trajectory reconstruction to both individual peak-age estimates and population-level covariate effects. Simulation studies show that, as observation windows become less informative, the joint model preserves substantially smaller regression bias and reflects the information loss through wider uncertainty, whereas pseudo-peak and two-stage alternatives exhibit larger bias. An application to World Athletics ranking data illustrates how the joint model estimates discipline-specific peak-age patterns and covariate effects under incomplete and irregular observation windows. Beyond sports, the proposed framework applies broadly to problems that couple longitudinal trajectories with function-derived latent features.
\end{abstract}


\section{Introduction}

\label{sec:intro}

Managing and forecasting athletic performance over the course of a career is a central problem in sport science. A particularly important target is the age of peak performance (APP), because the timing of peak ability informs talent identification, training design, career planning, and health management \citep{allen2015age}. In high-performance sport, these questions also arise within broader development systems involving public investment, family costs, dual-career pressures, athlete transitions, and well-being concerns \citep{cakravastia2022role,hardy2010parental,oydna2025commitment,pilkington2024athlete,spence2012nonrefundable,tessitore2021parents,houltberg2020developmental,soares2023burnout,subotnik2019environmental}. Our motivating application uses track-and-field data, where events differ substantially in physiological and technical demands and therefore provide a useful setting for studying discipline-specific peak-age patterns. Empirical studies across sports suggest that peak performance often occurs in the mid-to-late twenties, but with substantial variation across disciplines \citep{balatoni2020there,baker2017advances}.

Statistical inference on APP is difficult because the relevant object is not directly observed. Conceptually, each athlete has an underlying age-specific performance capacity, denoted by $Y_i(t)$ for athlete $i$ at age $t$. The latent peak age is
\[
\tau_i = \arg\max_t Y_i(t),
\]
the age at which the underlying performance trajectory reaches its maximum. In practice, however, the full trajectory $Y_i(t)$ is unavailable. We observe only noisy competition records
\[
y_{ij} = Y_i(t_{ij}) + \varepsilon_{ij},
\]
at irregular ages $t_{ij}$, often over incomplete observation windows. Here $\varepsilon_{ij}$ represents residual variation between observed competition performance and latent performance capacity, including short-term performance fluctuations and measurement noise. Because competition records are noisy, sparse, and irregularly timed, the age of the best recorded performance may differ from the latent APP of scientific interest. Thus, even when the scientific question is simple to state, the data provide only a fragmented view of the career trajectory. These same features also complicate regression analysis of how athlete-level characteristics shape both performance trajectories and peak age.

Most applied work in sport science therefore uses simplified summaries. One common approach defines a pseudo peak age, such as the age of the best observed performance or the average age of top-ranked athletes \citep{stiefel2013age,gorzi2022prediction,nikolaidis2017performance}. Such summaries are intuitive and easy to compute, but they can be biased or unstable when records are sparse, or observed over only part of a career. A second approach first fits individual performance trajectories and then regresses the resulting estimated peak ages on athlete-level covariates \citep{huebner2019performance,solberg2019peak,haugen2018peak,chen2025peak}. Recent mixed-effects and Bayesian longitudinal models improve trajectory estimation by borrowing information across athletes \citep{griffin2022bayesian,romann2024longitudinal,spyropoulou2024modelling}, but peak age is often still extracted as a fitted-curve summary. In such two-stage analyses, the estimated peak ages are treated as observed outcomes in the second stage, which can ignore first-stage uncertainty and lead to understated uncertainty in regression inference \citep{murphy2002estimation}. Incomplete observation windows can also induce censoring- or truncation-like behavior, and heuristic treatments, such as excluding athletes with few records or declaring the last observed age to be the peak, may introduce additional bias \citep{brander2014estimating}.

These limitations matter because APP is plausibly associated with multiple athlete-level and contextual factors. Athlete development pathways, maturation timing, relative age effects, early career milestones, specialization, training history, training load, recovery, and injury history may all shape long-term performance dynamics \citep{parry2024associations,barreira2021relative,li2020predicting,mclellan2022youth,exel2024precision,mian2023relationship}. This creates the need for a statistical framework that can keep the trajectory, the peak age, and covariate effects in the same inferential system.

This paper develops a Bayesian joint modeling framework for longitudinal performance records and latent peak ages. The proposed framework makes three main contributions. First, it reconstructs athlete-specific latent performance trajectories from noisy, irregular, and partially observed competition records. Second, it treats APP as a latent response defined by these trajectories, rather than as an observed best-performance age or a plug-in estimate from a first-stage curve fit. 
Third, it links latent peak age to athlete- and sport-specific covariates while propagating uncertainty from trajectory reconstruction and peak-age estimation into regression inference. 
In simulations that vary the informativeness of the observed age window and the trajectory specification, the proposed joint formulation yields less biased covariate-effect estimates and reflects reduced peak-age information through wider uncertainty intervals, whereas conventional alternatives tend to translate the same information loss into attenuation bias. Together, these components provide a coherent framework for estimating individual APPs, comparing peak-age patterns across sports and groups, and quantifying the uncertainty in those comparisons. An accompanying \textsf{R} package implements the proposed method.

This formulation also has a natural connection to several areas of statistical methodology. In failure-time settings, an event time may be right-, left-, or interval-censored \citep{klein2003survival}, and a large literature has developed regression methods for censored or interval-censored outcomes \citep{turnbull1976empirical,finkelstein1986proportional,tian2006accelerated,zhang2010interval}. Standard joint models link longitudinal biomarkers to externally observed clinical event times, such as death, disease progression, or injury \citep{wulfsohn1997joint,wu2008joint,brilleman2019joint,liu2026bayesian}, and related ideas have also appeared in sports settings involving longitudinal measurements and time-to-injury outcomes \citep{nielsen2019time}; see \cite{wang2025joint} for a recent overview. In contrast, APP is not an externally occurring event time observed through a censoring mechanism. It is a latent functional, $\tau_i=\arg\max_t Y_i(t)$, of an unobserved performance trajectory. The proposed framework also shares features with multi-path or hierarchical change-point models \citep{joseph1992estimation}, but the target here is not a structural breakpoint; it is the timing of optimal performance defined by a smooth latent trajectory.

The remainder of the paper is organized as follows. Section~\ref{sec:motivating-data} introduces the World Athletics data example that motivates the proposed framework. Section~\ref{sec:method} presents the joint model, including trajectory specifications, the peak-age submodel, and Bayesian inference. Section~\ref{sec:data-analysis} applies the model to the athletics data, Section~\ref{sec:simulation} reports simulation studies comparing the proposed method with conventional approaches, and Section~\ref{sec:discussion} concludes with limitations and future directions.

\section{Motivating Athletics Data Example}
\label{sec:motivating-data}

Before introducing the model, we first describe the athletics data example that motivates the inferential problem. The purpose of this example is not only to provide the empirical setting for the application, but also to make concrete the challenges described in the Introduction: performance is observed at irregular competition ages, observation windows are often incomplete, and the age of the best recorded performance need not coincide with the latent peak age of scientific interest.

The motivating dataset consists of elite athletics performance records obtained from the World Athletics all-time rankings database. The original records are publicly available from the official World Athletics website\footnote{\url{https://worldathletics.org/records/all-time-toplists}}, which compiles historical results from international competitions across events. To facilitate data access and reproducibility, we used a curated dataset available on Kaggle\footnote{\url{https://www.kaggle.com/datasets/jeannicolasduval/world-athletics-all-time-rankings}}, which was constructed by programmatically collecting the official rankings pages. Since the dataset is a direct scrape of the World Athletics database without modification of the underlying records, it provides a convenient and reproducible snapshot of the official data source used in this study. Because all-time ranking lists include only performances that meet the ranking criteria, the observed records primarily consist of high-level performances and may not capture every competition in an athlete's career.

The dataset contains athlete-level performance records across multiple athletics disciplines. Each entry corresponds to a competition result and includes the athlete name, date of birth, event, performance score, and competition date. Performance in athletics events is measured using the World Athletics scoring tables, which convert raw results, such as time or distance, into a standardized performance score that allows comparisons across events \citep{spiriev2000iaaf}. These scoring tables are widely used in multi-event competitions such as the decathlon and heptathlon to aggregate performances across disciplines.
For each athlete, multiple observations are typically available across different ages, forming an irregular longitudinal trajectory over the observed record window of their career. 
To capture heterogeneity in age-related performance development across athletics disciplines, we selected ten representative events spanning substantially different physiological and technical demands: 100 metres, 400 metres, 1500 metres, marathon, long jump, shot put, discus throw, javelin throw, decathlon (men), and heptathlon (women). To facilitate interpretation, these events are further grouped into six broader categories reflecting similar physiological and technical characteristics: sprints (100 m and 400 m), middle-long distance (1500 m), road running (marathon), jumps (long jump), throws (shot put, discus throw, and javelin throw), and combined events (decathlon (men) and heptathlon (women)). Both men's and women's records were included when available, with decathlon and heptathlon treated as sex-specific combined events.

We applied minimal preprocessing to construct athlete-level longitudinal records for trajectory modeling. Athlete identifiers were based on name and date of birth; duplicate records and two likely linkage errors were removed. Athletes without a recorded date of birth were assumed to be missing completely at random and excluded ($0.03\%$ of the dataset), and July~1 was used when only birth year was available. We retained athletes with at least ten recorded competitions and focused on those appearing in a single selected event. Among athletes meeting the ten-record threshold, 366 appeared in multiple selected events and were excluded. Such athletes raise important questions about event-specific trajectories, within-athlete dependence, and event transitions, which we leave to future work. After preprocessing, the analysis dataset contains 105{,}515 records from 3{,}229 single-event athletes, with one longitudinal trajectory per athlete; event-level summaries are provided in Supplementary Table~S1.

Figure~\ref{fig1} gives a more direct view of these challenges at the athlete level. The selected records show that observed performances can be sparse and irregular over age, and that the best observed performance may occur near the beginning or end of the observed record window. These patterns visually reinforce the distinction made in the Introduction between an observed or pseudo peak age and the latent peak age defined by the underlying performance trajectory.

\begin{figure}[!ht]
    \centering
    \includegraphics[width=1\linewidth]{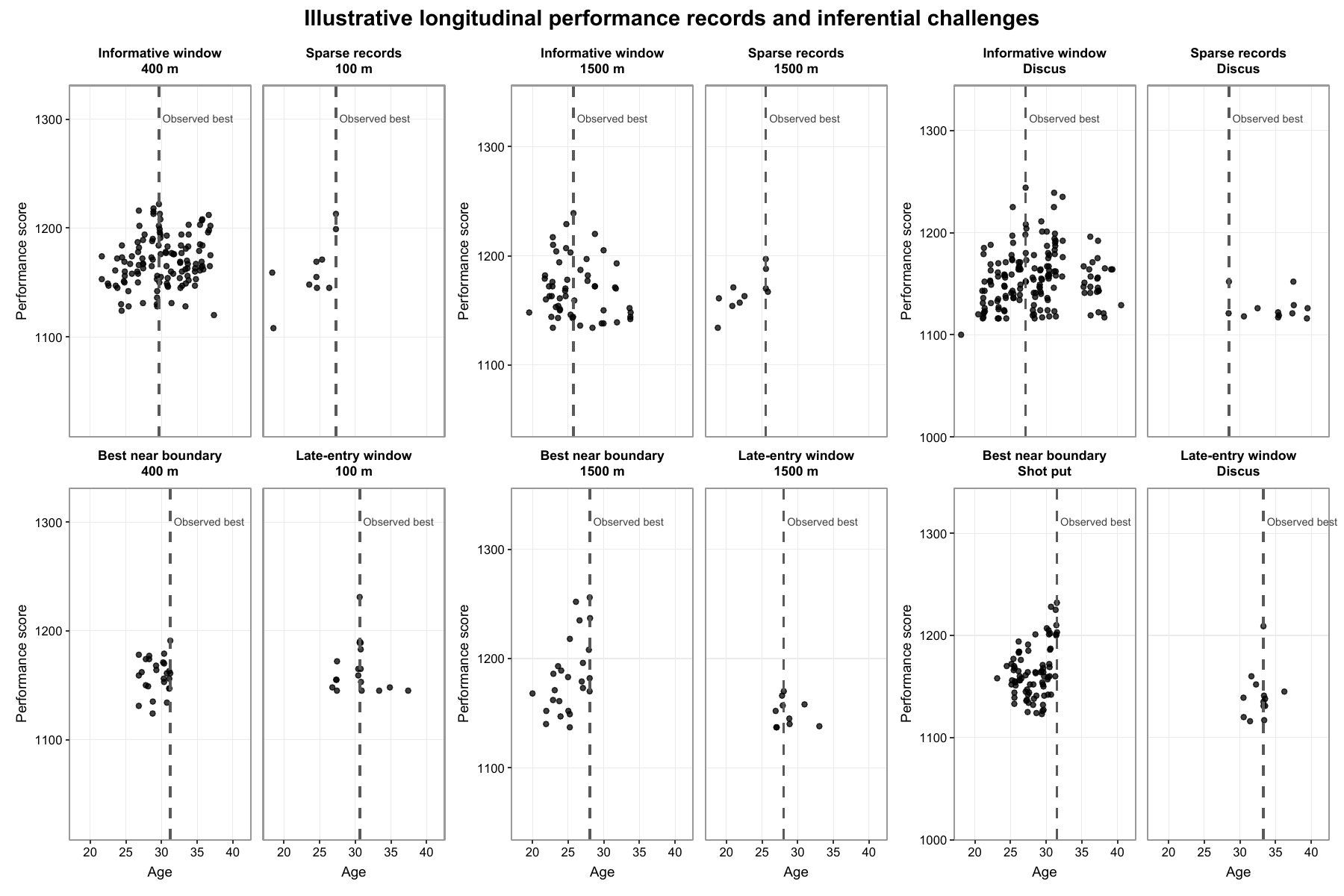}
    \caption{Observed performance records for selected elite male athletes across event types. Each panel shows one athlete. The examples illustrate sparse or irregular records, incomplete age windows, and best recorded performances near the boundary of the observed age range. Dashed vertical lines mark the age of the best recorded performance, which may differ from the latent peak age.}
    \label{fig1}
\end{figure}

Taken together, the World Athletics records provide a rich but incomplete longitudinal view of athlete development over age. They motivate a model that can borrow information across athletes and sports, reconstruct smooth athlete-specific trajectories from irregular observations, and treat peak age as a latent quantity rather than as a directly observed competition age. We now formalize this modeling problem.

\section{A Joint Modeling Framework}
\label{sec:method}
\subsection{Notation and Modeling}
\label{sec:method:1}
Let $i=1,\ldots,I$ index athletes and $j=1,\ldots,n_i$ index observed performance outcomes. Denote by $y_{ij}$ the observed performance score at age $t_{ij}$. We let $\mathbf{x}_i$ denote athlete-level covariates (e.g., biological sex, sport, birth cohort) and $\mathbf{z}_{ij}$ denote competition-level covariates such as wind speed, when available. Competition-level covariates $\mathbf{z}_{ij}$ adjust performance at a specific competition, whereas athlete-level covariates $\mathbf{x}_i$ enter the hierarchical distributions of athlete-specific trajectory parameters and latent peak age.

We model the longitudinal component of performance using a subject-specific smooth trajectory with additive residual variation. In the notation introduced in Section \ref{sec:intro}, this latent age-performance trajectory is parameterized as
$Y_i(t)=\mu(t;\boldsymbol{\theta}_i,\tau_i)$. Conditional on latent trajectory parameters and covariates, observed performances are assumed independent across competitions. Specifically,
\begin{equation}
y_{ij} = \mu(t_{ij};\boldsymbol{\theta}_i,\tau_i) 
+ \boldsymbol{\delta}^\top \mathbf{z}_{ij} 
+ \varepsilon_{ij},
\qquad 
\varepsilon_{ij} \sim N(0,\sigma_y^2),
\label{model:y}
\end{equation}

In \eqref{model:y}, $\varepsilon_{ij}$ denotes the remaining residual variation after adjusting for observed competition-level covariates through $\boldsymbol{\delta}^\top\mathbf{z}_{ij}$. $\mu(\cdot)$ is a prespecified trajectory family, the random vector
$\boldsymbol{\theta}_i$ governs the shape of the individual performance curve (e.g., growth and decline rates), 
and $\tau_i$ represents the age at which the trajectory attains its maximum. 
The vector $\boldsymbol{\delta}$ captures fixed effects of competition-level covariates $\mathbf{z}_{ij}$.

Between-athlete heterogeneity is modeled hierarchically through
\begin{equation}
\boldsymbol{\theta}_i \mid \mathbf{x}_i
\sim
F_\theta(\cdot \mid \mathbf{x}_i;\boldsymbol{\psi}_\theta),
\label{model:theta}
\end{equation}
where $F_\theta$ is a conditional distribution for the athlete-specific trajectory
parameters and $\boldsymbol{\psi}_\theta$ collects the corresponding population-level
parameters.  This notation is intended to keep the framework modular: the components of
$\boldsymbol{\theta}_i$ and the distribution $F_\theta$ may be chosen to match a
particular trajectory parameterization. For example, a normal random-effects
specification is
\[
\boldsymbol{\theta}_i = \boldsymbol{\theta}_0 
+ \mathbf{B}\mathbf{x}_i 
+ \boldsymbol{b}_i,
\qquad 
\boldsymbol{b}_i \sim \mathrm{MVN}(\mathbf{0},\boldsymbol{\Sigma}_\theta),
\]
where $\boldsymbol{\theta}_0$ represents baseline trajectory parameters, 
$\mathbf{B}$ links athlete-level covariates $\mathbf{x}_i$ to trajectory characteristics, 
and $\boldsymbol{b}_i$ captures residual subject-specific deviations. Under this
normal special case, $\boldsymbol{\psi}_\theta=(\boldsymbol{\theta}_0,\mathbf{B},
\boldsymbol{\Sigma}_\theta)$, and the covariance structure of
$\boldsymbol{\Sigma}_\theta$ can be specified to reflect dependence among the
components of $\boldsymbol{\theta}_i$.

A central scientific objective of this study is to understand how athlete-level
characteristics are associated with earlier or later peak performance development.
To quantify covariate effects on peak age, we specify a parametric submodel for
the latent peak age $\tau_i$. This submodel is embedded within the joint
longitudinal framework so that inference on $\tau_i$ and its covariate effects is
based on the full observed performance trajectory, rather than on the empirically
observed best performance alone. This distinction is especially important when
the observed window provides only partial information about the location of the
peak.
In particular, we adopt an accelerated failure time (AFT) formulation
\begin{equation}
\log(\tau_i) = \beta_0 + \boldsymbol{\beta}^\top \mathbf{x}_i + e_i,
\qquad e_i \sim N(0,\sigma_\tau^2),
\label{model:tau}
\end{equation}
so that $\tau_i \mid \mathbf{x}_i$ follows a lognormal distribution. Here, $\beta_0$ represents a baseline peak age on the log scale, while $\boldsymbol{\beta}$ captures how athlete-level covariates shift peak age multiplicatively. The variance parameter $\sigma_\tau^2$ reflects residual heterogeneity in peak age that is not explained by observed covariates. Notably, although we focus on the lognormal AFT model for concreteness, alternative parametric specifications such as proportional hazard models can be incorporated within the same joint framework.

\subsection{Specification of the performance trajectory function}
\label{sec:method:2}

Because peak age is defined as the maximizer of an athlete-specific performance trajectory, we use parametric longitudinal curves in which peak age can be represented directly as a trajectory parameter. This provides an interpretable way to infer $\tau_i$ from irregular and noisy performance records and to link it to athlete-level covariates. The proposed framework is modular with respect to the choice of the trajectory function $\mu(\cdot)$.
We consider two parametric specifications that differ in flexibility and interpretability.

\paragraph{Symmetric quadratic model}
A simple and interpretable specification assumes a symmetric quadratic trajectory,
\begin{equation}
\mu(t;\alpha,\rho,\tau) = \alpha - \rho (t-\tau)^2,
\label{eq:quad-sym}
\end{equation}
where $\alpha$ represents the peak performance level, $\rho>0$ controls the curvature of the trajectory and implies how rapidly performance changes with age, and $\tau$ denotes the peak age. 
This formulation is widely used in sports science (e.g. \cite{guan2024acceleration}) due to its parsimony, interpretability, and empirical adequacy in many applications. 
In particular, the quadratic form provides a convenient summary of peak level, peak age, and overall curvature within a low-dimensional parameterization. 
However, it imposes symmetry around the peak, implicitly assuming that the rate of performance improvement before the peak mirrors the rate of decline afterward. 
Such symmetry may not hold in sports where developmental progression and post-peak aging follow different dynamics.

\paragraph{Asymmetric quadratic model}
To allow distinct rates of improvement and decline, we extend the quadratic form to
\begin{equation}
\mu(t;\alpha,\rho_L,\rho_R,\tau) =
\begin{cases}
\alpha - \rho_L(\tau-t)^2, & t \le \tau,\\
\alpha - \rho_R(t-\tau)^2, & t > \tau,
\end{cases}
\label{eq:quad-asym}
\end{equation}
where $\rho_L$ and $\rho_R$ are positive parameters that govern the pre-peak and post-peak curvature, respectively. 
This specification accommodates asymmetric developmental patterns commonly observed in athletic performance.
Although the trajectory is defined piecewise, it remains continuous and differentiable at the peak age $\tau$, while allowing the curvature to differ before and after the peak.
For example, in sprint events, athletes may experience rapid performance improvement during early developmental years, whereas performance decline after the peak often occurs more gradually. In the notation of the general model, the latent trajectory parameter vector is
$\boldsymbol{\theta}_i=(\alpha_i,\rho_i)^\top$ under the symmetric quadratic
specification and $\boldsymbol{\theta}_i=(\alpha_i,\rho_{Li},\rho_{Ri})^\top$
under the asymmetric quadratic specification, with $\tau_i$ as the latent peak age in both cases.

The parameters $\alpha_i$, $\rho_{Li}$, $\rho_{Ri}$, and $\tau_i$ provide interpretable athlete-specific summaries of peak level, pre-peak development, post-peak decline, and peak age, respectively. These quantities may vary systematically across sports, since sprint, endurance, and technical events can involve different developmental patterns, peak ages, and aging dynamics. This consideration will guide the application-specific model specification in later sections, where sport-specific effects are incorporated into multiple components of the joint model.

\subsection{Joint likelihood formulation}

We now make explicit the likelihood structure implied by the proposed model. 
For athlete $i$, let $\mathbf{y}_i=(y_{i1},\ldots,y_{in_i})^\top$ denote the observed longitudinal outcomes, and 
\(\mathbf{z}_i=(\mathbf{z}_{i1},\ldots,\mathbf{z}_{in_i})\) 
denote the corresponding competition-level covariates.
The conditional density of $\mathbf{y}_i$, given the latent trajectory parameters $\boldsymbol{\theta}_i$, latent peak age $\tau_i$, and covariates, can be written as
\begin{equation}
f_y(\mathbf{y}_i 
\mid 
\boldsymbol{\theta}_i,\tau_i,
\mathbf{z}_i;
\boldsymbol{\delta},\sigma_y)
=
(2\pi\sigma_y^2)^{-n_i/2}
\exp\left\{
-\frac{1}{2\sigma_y^2}
\sum_{j=1}^{n_i}
\big[y_{ij}-\mu(t_{ij};\boldsymbol{\theta}_i,\tau_i)-\boldsymbol{\delta}^\top\mathbf{z}_{ij}\big]^2
\right\}.
\label{eq:fy}
\end{equation}

The latent trajectory parameters $(\boldsymbol{\theta}_i,\tau_i)$ are assumed independent across athletes conditional on covariates and global parameters $(\boldsymbol{\psi}_\theta,
\beta_0,\boldsymbol{\beta},\sigma_\tau)$. Given athlete-level covariates \(\mathbf x_i\) and shared model parameters, let \(f_\theta\) and \(f_\tau\) denote the conditional densities of the latent trajectory parameters and latent peak age, respectively. Under the normal random-effects specification in \eqref{model:theta} and the lognormal AFT specification in \eqref{model:tau},
\begin{align}
f_{\theta}(\boldsymbol{\theta}_i
\mid \mathbf{x}_i;
\boldsymbol{\psi}_\theta)
&=
\phi_{\mathrm{MVN}}\!\left(
\boldsymbol{\theta}_i;\,
\boldsymbol{\theta}_0+\mathbf{B}\mathbf{x}_i,\,
\boldsymbol{\Sigma}_\theta
\right),
\label{eq:ftheta} \\
f_{\tau}(\tau_i
\mid \mathbf{x}_i;
\beta_0,\boldsymbol{\beta},\sigma_\tau)
&=
\frac{1}{\tau_i\sigma_\tau\sqrt{2\pi}}
\exp\left\{
-\frac{(\log\tau_i-\beta_0-\boldsymbol{\beta}^\top\mathbf{x}_i)^2}
{2\sigma_\tau^2}
\right\}.
\label{eq:ftau}
\end{align}
 More generally, $f_\theta$ denotes the density induced by
the chosen conditional distribution $F_\theta$ in \eqref{model:theta}. In
\eqref{eq:ftheta}, $\phi_{\mathrm{MVN}}(\cdot;\boldsymbol{\mu},\boldsymbol{\Sigma})$ 
denotes the multivariate normal density with mean $\boldsymbol{\mu}$ and
covariance matrix $\boldsymbol{\Sigma}$.

Let $\boldsymbol{\Theta}
=
(\boldsymbol{\delta}, \sigma_y,
\boldsymbol{\psi}_\theta,
\beta_0,\boldsymbol{\beta},\sigma_\tau)$ collect all unknown global parameters. Under a concrete choice of $F_\theta$, $\boldsymbol{\psi}_\theta$ is replaced by the
corresponding distributional parameters, such as $(\boldsymbol{\theta}_0,\mathbf{B},
\boldsymbol{\Sigma}_\theta)$ in the normal special case. We assume that, conditional on covariates and $\boldsymbol{\Theta}$, the latent trajectory parameters are independent across athletes according to $f_\theta$, the AFT errors $\{e_i\}_{i=1}^I$ are independent across athletes according to \eqref{model:tau}, and the measurement errors $\{\varepsilon_{ij}: i=1,\ldots,I;\; j=1,\ldots,n_i\}$ are independent across observations; the trajectory and peak-age components are also conditionally independent given the covariates and shared parameters. If the latent variables $(\boldsymbol{\theta}_i,\tau_i)$ were observed, the complete-data likelihood contribution for athlete $i$ would be
\begin{equation}
L_i^{(c)}(\boldsymbol{\Theta}
\mid
\mathbf{y}_i,\boldsymbol{\theta}_i,\tau_i,
\mathbf{x}_i,\mathbf{z}_i)
=
f_y(\mathbf{y}_i 
\mid 
\boldsymbol{\theta}_i,\tau_i,
\mathbf{z}_i;
\boldsymbol{\delta},\sigma_y)
\,
f_{\theta}(\boldsymbol{\theta}_i
\mid \mathbf{x}_i;
\boldsymbol{\psi}_\theta)\,
f_{\tau}(\tau_i
\mid \mathbf{x}_i;
\beta_0,\boldsymbol{\beta},\sigma_\tau).
\label{eq:complete-lik}
\end{equation}

However, the latent quantities $(\boldsymbol{\theta}_i,\tau_i)$ are not directly observed in practice.
Assuming independence across athletes, the observed-data likelihood based on
$\mathcal{D} = \{(y_{ij}, t_{ij}, \mathbf{z}_{ij}, \mathbf{x}_i):
i=1,\ldots,I;\; j=1,\ldots,n_i\}$ is therefore obtained by integrating out the latent trajectory parameters and latent peak ages. 
For notational simplicity, we suppress the dependence of the component densities on shared model parameters in the following expression.
\begin{equation}
L^{(o)}(\boldsymbol{\Theta}\mid\mathcal{D})
=
\prod_{i=1}^{I}\left[
\int_{\mathcal{S}_\theta} \int_{0}^{\infty}
f_y(\mathbf{y}_i \mid \boldsymbol{\theta}_i,\tau_i,\mathbf{z}_i)
\, 
f_{\theta}(\boldsymbol{\theta}_i\mid\mathbf{x}_i)
\,
f_{\tau}(\tau_i\mid\mathbf{x}_i)
\,
d\tau_i\, d\boldsymbol{\theta}_i
\right],
\label{eq:obs-lik}
\end{equation}
where $\mathcal{S}_\theta$ denotes the support of the athlete-specific latent trajectory parameters.
This likelihood is not a standard survival likelihood for directly censored or truncated event times. Instead, incomplete observation windows affect inference through the longitudinal component $f_y$, and uncertainty about $\tau_i$ is propagated by integrating over the latent trajectory parameters and latent peak age.

In general, the integrals in \eqref{eq:obs-lik} do not admit closed-form expressions. 
We therefore adopt a Bayesian estimation framework, which provides a coherent approach for inference in hierarchical latent-variable models and yields full uncertainty quantification for both trajectory parameters and peak ages. Bayesian methods are increasingly used in sports analytics, particularly in settings involving multi-level performance data, partial career records, and complex dependency structures. In addition, sports science applications often benefit from incorporating substantive prior knowledge regarding performance development, physiological constraints, or sport-specific characteristics. Some recent examples include \cite{peng2024bayesian,tea2023analysis, peng2022bayesian,d2023bayesian,dolmeta2023bayesian,hecksteden2022dwarfs}.

\subsection{Simultaneous Bayesian inference for parameter estimation}

From a Bayesian perspective, the proposed framework constitutes a hierarchical
latent-variable model that jointly estimates population-level parameters,
athlete-specific trajectory parameters, and latent peak ages. This avoids a
two-stage analysis in which individual peak ages are first estimated from fitted
performance curves and then treated as observed outcomes. The hierarchy naturally reflects the data-generating mechanism described in the previous section. At the first level, the longitudinal outcomes \(\mathbf{y}_i\) are modeled conditionally on athlete-specific latent trajectory parameters, latent peak age, competition-level covariates, and observation-model parameters. At the second level, the latent trajectory parameters are modeled through regression structures on covariates $\mathbf{x}_i$, inducing trajectory-distribution parameters $\boldsymbol{\psi}_\theta$ (for example, $(\boldsymbol{\theta}_0,\mathbf{B},\boldsymbol{\Sigma}_\theta)$ under the normal special case) for $\boldsymbol{\theta}_i$ and $(\beta_0,\boldsymbol{\beta},\sigma_\tau)$ for $\tau_i$. 

Specifically, the hierarchical structure can be summarized in compact form as
\[
\mathbf{y}_i \mid \boldsymbol{\theta}_i,\tau_i,\mathbf{z}_i,\boldsymbol{\delta},\sigma_y 
\sim f_y(\cdot),
\]
\[
\boldsymbol{\theta}_i \mid \mathbf{x}_i 
\sim f_{\theta}(\cdot), 
\qquad
\tau_i \mid \mathbf{x}_i 
\sim f_{\tau}(\cdot),
\]
as defined in \eqref{eq:fy}--\eqref{eq:ftau}. 

Under the Bayesian framework, inference proceeds by combining the likelihood with prior distributions on the global parameters $\boldsymbol{\Theta}$. Let $\pi(\boldsymbol{\Theta})$ denote the joint prior density. The joint posterior distribution of all unknown quantities, including the latent variables, is proportional to
\begin{equation}
\pi\!\left(
\boldsymbol{\Theta},
\{\boldsymbol{\theta}_i,\tau_i\}_{i=1}^I
\mid
\mathcal{D}
\right)
\propto
\left[
\prod_{i=1}^{I}
L_i^{(c)}(\boldsymbol{\Theta}
\mid
\mathbf{y}_i,\boldsymbol{\theta}_i,\tau_i,
\mathbf{x}_i,\mathbf{z}_i)
\right]
\,
\pi(\boldsymbol{\Theta}),
\label{eq:joint-posterior}
\end{equation}
where $L_i^{(c)}(\boldsymbol{\Theta}\mid\cdot)$ is the complete-data likelihood defined in \eqref{eq:complete-lik}. 

Marginal posterior inference for $\boldsymbol{\Theta}$ is obtained by integrating out the latent variables,
\[
\pi(\boldsymbol{\Theta}\mid\mathcal{D})
=
\int
\pi\!\left(
\boldsymbol{\Theta},
\{\boldsymbol{\theta}_i,\tau_i\}_{i=1}^I
\mid
\mathcal{D}
\right)
\,
\prod_{i=1}^I d\boldsymbol{\theta}_i\, d\tau_i,
\]
which mirrors the observed-data likelihood formulation in \eqref{eq:obs-lik}. In practice, direct evaluation of these high-dimensional integrals is computationally infeasible. We therefore draw samples from the joint posterior \eqref{eq:joint-posterior} using Markov chain Monte Carlo (MCMC) methods, treating the latent variables $(\boldsymbol{\theta}_i,\tau_i)$ as additional unknowns.

An important advantage of this simultaneous Bayesian inference is that it provides direct posterior inference for the latent peak ages. In particular, for each athlete, the posterior distribution \(\pi(\tau_i\mid\mathcal{D})\) summarizes both the estimated peak age and its uncertainty, rather than reducing the athlete's longitudinal record to a single estimated peak age. This is especially useful for athletes with sparse, noisy, or incomplete performance records, including cases in which the true peak may lie outside the observed age range. The resulting posterior uncertainty for \(\tau_i\) is propagated naturally into inference on population-level peak-age effects, such as the regression coefficients \(\boldsymbol{\beta}\).

In practical implementation, a joint prior distribution is assigned to the global parameters $\boldsymbol{\Theta}$ to complete the Bayesian specification. Since the proposed framework is intended to accommodate a range of trajectory specifications and application contexts, we adopt weakly informative priors reflecting plausible ranges of athletic performance characteristics while maintaining sufficient flexibility for the data to dominate inference. 

The specific prior choices and computational implementation used for the empirical analysis are described in Section~\ref{sec:data-analysis}, where we apply a particular specification of the proposed framework to the motivating athletics data introduced in Section~\ref{sec:motivating-data}.

\section{Application to World Athletics Data}
\label{sec:data-analysis}

\subsection{Model and prior specifications for the application}

Using the World Athletics data described in Section~\ref{sec:motivating-data},
the model used in this application is a specific instantiation of the general
framework introduced in Section~\ref{sec:method}, with a parameterization
tailored to the structure of athletics performance data.

In this application, the athlete-level covariates include sport category, biological sex, birth year, and, for the peak-age submodel, sport-by-sex interactions. Because sprints, middle-long distance, road running, jumps, throws, and combined events may differ in peak performance levels, developmental rates, post-peak decline, and peak age, we allow sport-category effects to enter multiple components of the model. In the notation of the general framework, this application uses the asymmetric quadratic trajectory in \eqref{eq:quad-asym} and takes the athlete-specific trajectory parameter vector to be
$\boldsymbol{\theta}_i=(\alpha_i,\log\rho_{Li},\log\rho_{Ri})^\top$,
corresponding to peak level, log pre-peak curvature, and log post-peak curvature,
respectively. Thus, the model below is a concrete special case of
\eqref{model:y}, \eqref{model:theta}, and \eqref{model:tau}, with $F_\theta$
specified through the random effects for $\alpha_i$, $\log\rho_{Li}$, and
$\log\rho_{Ri}$.

For athlete $i$ and competition $j$, the observed performance score is modeled as
\begin{equation}
\label{eq:app-model}
\begin{aligned}
y_{ij}
&=
\mu(t_{ij};\alpha_i,\rho_{Li},\rho_{Ri},\tau_i) + \varepsilon_{ij},
\qquad
\varepsilon_{ij} {\sim} N(0,\sigma_y^2),
\\[4pt]
\mu(t;\alpha_i,\rho_{Li},\rho_{Ri},\tau_i)
&=
\begin{cases}
\alpha_i - \rho_{Li}(\tau_i-t)^2, & t \le \tau_i,\\
\alpha_i - \rho_{Ri}(t-\tau_i)^2, & t > \tau_i,
\end{cases}
\\[6pt]
\alpha_i
&=
\alpha_0 + \eta^{(\alpha)}_{\mathrm{sport}[i]} + u^{(\alpha)}_i,
\\[4pt]
\log \rho_{Li}
&=
\log \rho_{L0} + \eta^{(L)}_{\mathrm{sport}[i]} + u^{(L)}_i,
\\[4pt]
\log \rho_{Ri}
&=
\log \rho_{R0} + \eta^{(R)}_{\mathrm{sport}[i]} + u^{(R)}_i,
\\[4pt]
\log(\tau_i)
&=
\beta_0
+ \eta^{(\tau)}_{\mathrm{sport}[i]}
+ \beta_S\,\mathrm{Female}_i
+ \beta_Y\,\mathrm{BirthYear}_i
\\[-1pt]
&\quad
+ \eta^{(\tau S)}_{\mathrm{sport}[i]}\,\mathrm{Female}_i
+ e_i,
\\[4pt]
&\begin{array}{@{}r@{\;}c@{\;}l@{\quad}r@{\;}c@{\;}l@{}}
u^{(\alpha)}_i &\sim& N(0,\sigma_{\alpha}^2), &
u^{(L)}_i &\sim& N(0,\sigma_L^2),\\[-1pt]
u^{(R)}_i &\sim& N(0,\sigma_R^2), &
e_i &\sim& N(0,\sigma_\tau^2).
\end{array}
\end{aligned}
\end{equation}

As in the asymmetric trajectory specification in Section~\ref{sec:method}, $\alpha_i$ denotes latent peak performance, $\rho_{Li}$ and $\rho_{Ri}$ denote pre- and post-peak curvature, and $\tau_i$ denotes latent peak age.

The variable $\mathrm{Female}_i$ indicates female biological sex.

The random effects $u_i^{(\alpha)}$, $u_i^{(L)}$, and $u_i^{(R)}$ capture athlete-level heterogeneity in peak level and in the two log-curvature components; the log-scale curvature parameterization enforces positivity and improves numerical stability.
More generally, the four athlete-level deviations $(u_i^{(\alpha)},u_i^{(L)},u_i^{(R)},e_i)$ could be modeled jointly using a multivariate normal distribution to account for residual correlations among peak level, pre-peak curvature, post-peak curvature, and peak age.

The sport-category terms are modeled as fixed effects relative to a reference category. The six sport categories are sprints, middle-long distance, road running, jumps, throws, and combined events, with middle-long distance used as the reference category.
Sport category effects are incorporated through $\eta^{(\alpha)}_{\mathrm{sport}[i]}$, $\eta^{(L)}_{\mathrm{sport}[i]}$, and $\eta^{(R)}_{\mathrm{sport}[i]}$, while peak age is modeled through an accelerated failure time structure with covariates including sport category, biological sex, birth year, and sport-by-sex interactions. The terms $\eta^{(\alpha)}_{\mathrm{sport}[i]}$, $\eta^{(L)}_{\mathrm{sport}[i]}$, $\eta^{(R)}_{\mathrm{sport}[i]}$, and $\eta^{(\tau)}_{\mathrm{sport}[i]}$ are sport-category fixed effects, whereas $\eta^{(\tau S)}_{\mathrm{sport}[i]}$ represents sport-by-sex interaction effects in the peak-age submodel.

Competition-level covariates $\mathbf{z}_{ij}$ are not included in this application, since no such variables are consistently available across all events considered here. However, the general framework allows such covariates to be incorporated when appropriate data are available.

This hierarchical specification allows information to be shared across athletes within sport categories, helping stabilize inference for athletes with sparse or incomplete longitudinal records.
Weakly informative priors are assigned to all global parameters. The priors are chosen to encode broad plausible ranges while allowing the likelihood to dominate when the data are informative; for example, the baseline peak-age intercept is centered at $\log(28)$ with standard deviation $0.5$, reflecting a wide prior range for typical elite peak ages on the original age scale. Similarly, curvature parameters are assigned priors on the log scale to enforce positivity while allowing substantial variation in trajectory shape. The full prior specification is provided in the Supplementary Material.

\subsection{Model fit and reconstructed performance trajectories}

Posterior inference was carried out using MCMC implemented in \texttt{Stan}. We ran 4 parallel chains with 1000 warm-up iterations followed by 2000 sampling iterations per chain.
Table~\ref{tab:post-all} provides an overall summary of the posterior estimates
for the key model parameters, including the baseline trajectory parameters,
covariate effects, and variance components. 

\begin{table}[!ht]
\centering
\caption{Posterior means and 95\% credible intervals for key parameters in the application-specific model. Male middle-long distance athletes are the reference group. Parameters are grouped by peak level, curvature, peak age, and variance components.}
\label{tab:post-all}
\small
\setlength{\tabcolsep}{4pt}
\renewcommand{\arraystretch}{1.1}
\begin{tabular}{llllll}
\hline
\multicolumn{3}{c}{\textbf{Parameter}} & \multicolumn{3}{c}{\textbf{Parameter}} \\
\cline{1-3} \cline{4-6}
Parameter & Mean & 95\% CrI & Parameter & Mean & 95\% CrI \\
\hline
\multicolumn{6}{l}{\textit{Peak-level submodel}} \\
$\alpha_0$ & 1166.933 & (1164.720, 1169.132) & $\eta^{(\alpha)}_{\mathrm{combined}}$ & 0.823 & (-9.648, 11.405) \\
$\eta^{(\alpha)}_{\mathrm{sprint}}$ & 0.254 & (-2.591, 2.963) & $\eta^{(\alpha)}_{\mathrm{throws}}$ & -26.734 & (-29.535, -23.948) \\
$\eta^{(\alpha)}_{\mathrm{road}}$ & 13.010 & (8.556, 17.348) & $\eta^{(\alpha)}_{\mathrm{jumps}}$ & 3.893 & (-0.160, 8.018) \\
\hline
\multicolumn{3}{l}{\textit{Peak-age main effects}} & \multicolumn{3}{l}{\textit{Sport-by-sex interactions}} \\
$\beta_0$ & 3.185 & (3.146, 3.224) & $\eta^{(\tau S)}_{\mathrm{combined}}$ & -0.029 & (-0.154, 0.094) \\
$\beta_S$ & 0.082 & (0.049, 0.117) & $\eta^{(\tau S)}_{\mathrm{sprint}}$ & -0.044 & (-0.084, -0.003) \\
$\beta_Y$ & 0.049 & (0.014, 0.083) & $\eta^{(\tau S)}_{\mathrm{throws}}$ & -0.035 & (-0.076, 0.004) \\
$\eta^{(\tau)}_{\mathrm{combined}}$ & 0.004 & (-0.105, 0.106) & $\eta^{(\tau S)}_{\mathrm{road}}$ & -0.069 & (-0.155, 0.018) \\
$\eta^{(\tau)}_{\mathrm{sprint}}$ & -0.103 & (-0.138, -0.067) & $\eta^{(\tau S)}_{\mathrm{jumps}}$ & 0.013 & (-0.040, 0.067) \\
$\eta^{(\tau)}_{\mathrm{throws}}$ & 0.044 & (0.011, 0.079) &  &  &  \\
$\eta^{(\tau)}_{\mathrm{road}}$ & 0.085 & (0.009, 0.158) &  &  &  \\
$\eta^{(\tau)}_{\mathrm{jumps}}$ & -0.088 & (-0.135, -0.042) &  &  &  \\
\hline
\multicolumn{6}{l}{\textit{Curvature submodel}} \\
$\log\rho_{L0}$ & -0.951 & (-1.198, -0.715) & $\log\rho_{R0}$ & -2.704 & (-3.093, -2.335) \\
$\eta^{(L)}_{\mathrm{combined}}$ & 0.619 & (-0.026, 1.251) & $\eta^{(R)}_{\mathrm{combined}}$ & 0.187 & (-0.723, 1.055) \\
$\eta^{(L)}_{\mathrm{sprint}}$ & 1.098 & (0.820, 1.379) & $\eta^{(R)}_{\mathrm{sprint}}$ & -0.347 & (-0.765, 0.064) \\
$\eta^{(L)}_{\mathrm{throws}}$ & 0.719 & (0.456, 0.980) & $\eta^{(R)}_{\mathrm{throws}}$ & 1.032 & (0.650, 1.431) \\
$\eta^{(L)}_{\mathrm{road}}$ & -0.458 & (-1.058, 0.164) & $\eta^{(R)}_{\mathrm{road}}$ & -0.534 & (-1.164, 0.099) \\
$\eta^{(L)}_{\mathrm{jumps}}$ & 0.735 & (0.357, 1.112) & $\eta^{(R)}_{\mathrm{jumps}}$ & 0.347 & (-0.117, 0.817) \\
\hline
\multicolumn{6}{l}{\textit{Variance parameters}} \\
$\sigma_y$ & 24.922 & (24.814, 25.033) & $\sigma_\tau$ & 0.099 & (0.093, 0.105) \\
$\sigma_\alpha$ & 27.852 & (27.091, 28.609) & $\sigma_L$ & 0.804 & (0.737, 0.876) \\
$\sigma_R$ & 1.563 & (1.420, 1.719) &  &  &  \\
\hline
\end{tabular}
\end{table}

Overall, the posterior estimates are well identified and support meaningful
statistical inference on both trajectory dynamics and peak-age effects. In particular, the variance components associated with the athlete-specific random
effects are clearly non-negligible. The estimated standard deviations for the
random effects in peak level, pre-peak curvature, and post-peak curvature are $\sigma_\alpha = 27.9$ (95\% credible interval (CrI): 27.1, 28.6), $\sigma_L = 0.80$ (95\% CrI: 0.74, 0.88), and $\sigma_R = 1.56$ (95\% CrI: 1.42, 1.72), respectively,
all clearly different from zero. These results indicate substantial between-athlete
heterogeneity in both overall performance level and the rates of performance
evolution, and provide
strong empirical support for incorporating random effects to capture
individual-level variation in performance trajectories.

\begin{figure}[!ht]
    \centering
    \includegraphics[width=1\linewidth]{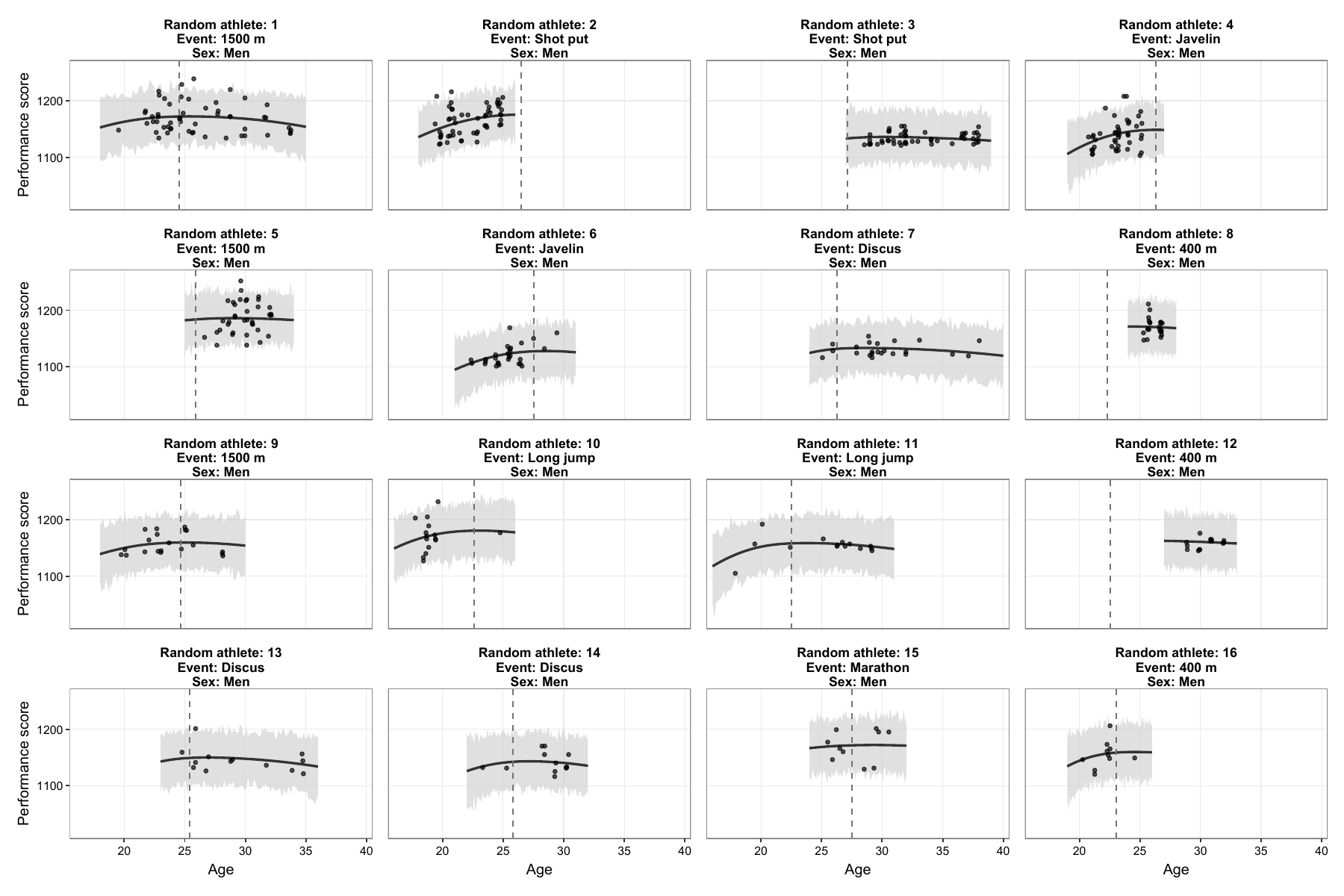}
    \caption{
Observed scores, posterior mean trajectories, and 95\% posterior predictive
intervals for 16 randomly selected male athletes. Solid curves show posterior
mean latent trajectories, shaded bands show posterior predictive intervals, and
dashed vertical lines mark posterior mean peak ages.
}
    \label{fig:trajectory_examples}
\end{figure}

To further illustrate model fit at the athlete level, Figure~\ref{fig:trajectory_examples}
displays observed performance scores together with the posterior mean trajectory
and the corresponding 95\% posterior predictive interval for a random sample of 16 male athletes across different disciplines. The vertical dashed line marks
the posterior mean estimate of the peak age $\tau_i$ for each athlete.

Across athletes with dense, sparse, or age-limited records, the fitted curves provide reasonable reconstructions of performance trajectories and plausible peak-age estimates.

\subsection{Covariate effects on peak age}

\begin{figure}[!ht]
    \centering
    \includegraphics[width=0.7\linewidth]{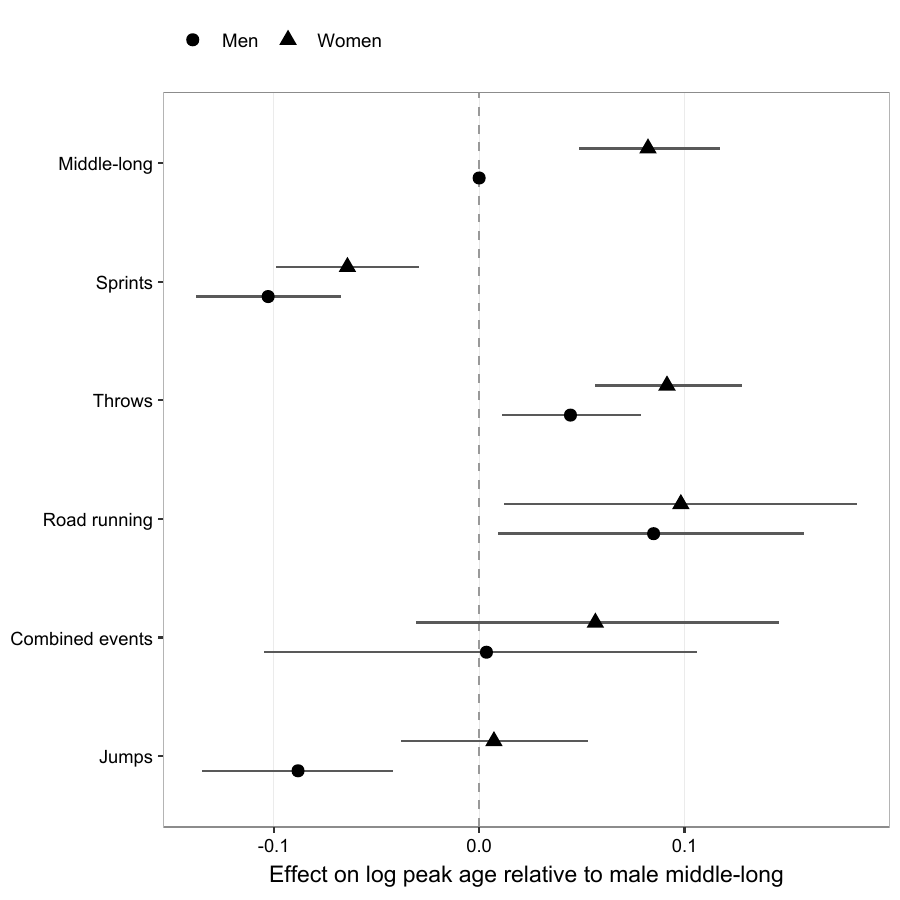}
    \caption{Posterior mean effects on log peak age from the accelerated failure time submodel by sport category and biological sex, relative to male middle-long distance athletes. Points and horizontal line segments show posterior means and 95\% credible intervals. Negative and positive values indicate earlier and later peak ages, respectively. The dashed vertical line marks zero effect.}
    \label{fig:peak_age_effects}
\end{figure}

The regression structure in the peak-age submodel (\ref{model:tau}) allows
direct statistical inference on factors associated with the timing of peak
performance. Table~\ref{tab:post-all} summarizes posterior estimates for all
model parameters, while Figure~\ref{fig:peak_age_effects} displays the posterior
means and 95\% credible intervals for sport-category and biological-sex effects on log peak age, expressed relative to the male middle-long distance reference group.

Figure~\ref{fig:peak_age_effects} shows two main patterns in the estimated sport-category and biological-sex effects on peak age.
First, estimated peak age varies systematically across sport
categories. Sprint and jumping events show clearly earlier peak ages compared
with middle-long distance events, while road running and throwing events tend to exhibit later peak ages, with the road-running effect most clearly separated from the middle-long reference group among men. These patterns are consistent
with physiological differences across disciplines, where explosive power events
typically favor earlier peak performance, whereas endurance- and technique-based
events allow athletes to remain competitive at older ages. 

Second, biological-sex differences in peak age are also evident. Across most sport
categories, female athletes tend to exhibit slightly later peak ages than male
athletes, although the magnitude of this difference varies by discipline. 

This pattern contrasts with the common expectation that female athletes peak
earlier due to earlier biological maturation. One possible explanation is that
the present analysis focuses on elite competition records, where peak
performance reflects not only physiological development but also career
dynamics such as selection, participation pathways, and competitive longevity.
For instance, female athletes may maintain elite-level competitiveness over
longer periods or achieve their highest-ranked performances at slightly older
ages due to differences in career progression and competitive structure.

In addition to the sport and biological-sex effects shown in the figure, the estimated birth-year coefficient is modestly positive ($\beta_Y=0.049$, 95\% CrI: $0.014$ to $0.083$). Birth year is rescaled to the unit interval using the minimum and maximum birth years in the analysis sample, so $\beta_Y$ compares athletes at the latest observed birth year with those at the earliest observed birth year rather than representing a one-year effect. On the original age scale, the posterior mean corresponds to a multiplicative effect of $e^{0.049}\approx1.050$, or about a 5.0\% higher peak age for athletes at the latest versus earliest observed birth year. This modest positive estimate may reflect possible cohort-related changes in training, competition opportunities, or athlete development pathways.

Supplementary Figure~S1 provides a complementary boxplot summary of athlete-level posterior mean peak ages on the original age scale, showing the distribution of estimated peak ages by sport category and biological sex.

\subsection{Sport differences in improvement and decline rates}

\begin{figure}[!ht]
    \centering
    \includegraphics[width=1\linewidth]{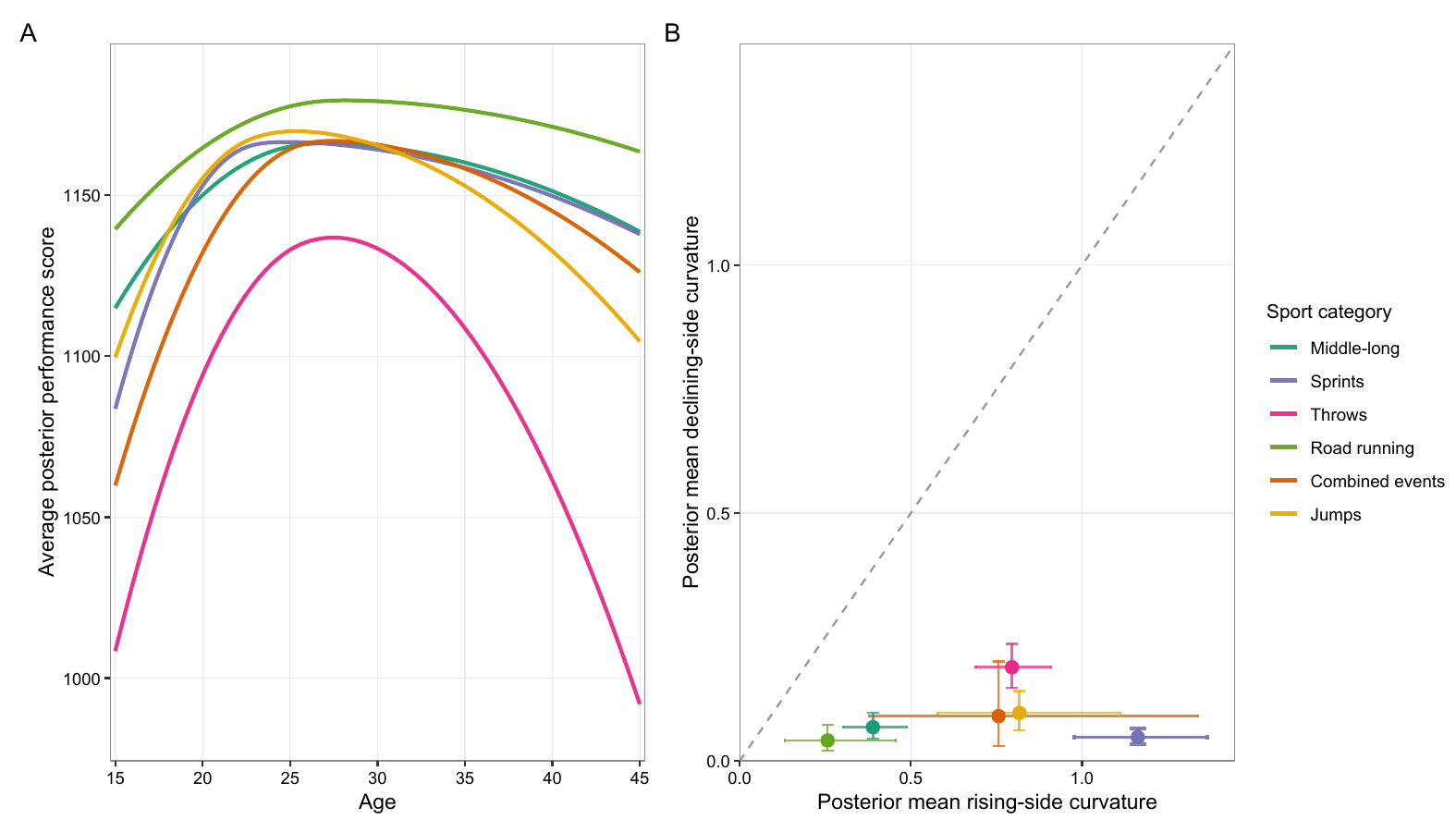}
    \caption{Sport-specific trajectory shapes. Panel A shows posterior mean average performance trajectories by sport category. Panel B shows posterior mean rising- and declining-side curvatures, $\rho_L$ and $\rho_R$; larger values indicate faster improvement or decline. Points and line segments show posterior means and 95\% credible intervals. The dashed diagonal marks $\rho_L=\rho_R$, corresponding to a symmetric quadratic trajectory.}
    \label{fig:curvature_effects}
\end{figure}

The proposed framework also enables comparisons of how performance evolves
before and after the peak across different sport categories. Posterior
summaries of the curvature-related parameters are reported in
Table~\ref{tab:post-all}, while Figure~\ref{fig:curvature_effects}
provides a visual summary of the sport-specific curvature estimates.

Figure~\ref{fig:curvature_effects}A displays the posterior mean average performance trajectory for each sport category, providing a direct visualization of how the fitted age-performance curves differ in peak level and shape across sports.
In Figure~\ref{fig:curvature_effects}B, the horizontal axis represents the rising-side curvature $\rho_L$, which governs the rate of performance improvement before the peak, and the vertical axis represents the declining-side curvature $\rho_R$, which governs the rate of performance decline after the peak. Larger curvature values correspond to more rapid changes in performance. The dashed line indicates equality between the two curvature parameters, corresponding to a symmetric quadratic trajectory.

Figure~\ref{fig:curvature_effects}A suggests that sprinting and jumping events have relatively similar average trajectory shapes, with rapid improvement before the peak followed by a more gradual post-peak decline. Figure~\ref{fig:curvature_effects}B clarifies that this pattern is most pronounced for sprints: sprinting events have the largest rising-side curvature and among the smallest declining-side curvature estimates. Jumping events also show relatively large rising-side curvature, but their declining-side curvature is higher than that of sprints, indicating somewhat sharper post-peak changes.

Middle-long distance and road-running events have smaller curvature parameters overall, especially road running, consistent with more gradual performance evolution and longer sustained periods of high-level performance. Throwing events have relatively high curvature on both sides, and the largest declining-side curvature among the sport categories, indicating sharper post-peak changes than in the other events. Combined events show greater uncertainty, reflecting their smaller sample size and higher variability.
In addition, the observed asymmetry between improvement and decline rates across
multiple sport categories provides empirical support for adopting a
non-symmetric performance trajectory specification.

\subsection{Model diagnostics and software implementation}

Convergence diagnostics for the MCMC algorithm, including trace plots and
comparisons between prior and posterior densities for key parameters, are
provided in Section~S4 of the Supplementary Material. These results indicate stable sampling
behavior and demonstrate that the data are informative for the major model
components. Additional checks on bias, uncertainty calibration, incomplete observation windows, trajectory misspecification, and individual peak-age estimation are provided in the simulation study in Section~\ref{sec:simulation}.

To facilitate reproducibility and future applications, the proposed methodology
has been implemented in an \textsf{R} package supporting both symmetric and
asymmetric trajectory specifications considered in this study. The package
provides functions for fitting the joint model, summarizing posterior peak-age
estimates, extracting athlete-level quantities, and visualizing fitted
performance trajectories. An anonymized version will be made available for
review, and the public repository will be released upon acceptance.

\section{Simulation study}
\label{sec:simulation}

\subsection{Simulation design}

Our simulation study is designed to mirror the main challenges in the empirical
application, including noisy longitudinal measurements, irregular observation
times, and incomplete observation windows. We focus on three aspects of performance:
(i) bias and variability of covariate effect estimates in the peak-age model;
(ii) the quality of uncertainty quantification, assessed by agreement between
Monte Carlo variability and reported standard errors or posterior uncertainty;
and (iii) sensitivity to incomplete observational windows, measured by how often
the true peak age $\tau_i$ lies within the observed age range. Here, Monte Carlo variability refers to the empirical variability of estimates across repeated simulated datasets, whereas reported standard errors or posterior standard deviations quantify the uncertainty reported by the method within each dataset. Agreement between these quantities provides a check on whether the method yields calibrated uncertainty estimates.

The data-generating procedure is summarized in Supplementary Algorithm~S1. In each setting, we generate datasets with $I=200$ athletes and summarize results over 200 replications. Peak ages are generated from a lognormal AFT model with true parameters $\beta_0=3.20$, $\beta_1=0.01$, $\beta_2=0.05$, and $\sigma_\tau=0.10$, while athlete-specific performance trajectories are generated under either a symmetric or asymmetric quadratic form. Unless otherwise stated, covariates, AFT errors, trajectory parameters, observation windows, and measurement errors are generated independently across athletes and observations. To assess sensitivity to incomplete observation, we consider three observation-window settings in which the true peak age falls within the observed age range with probability approximately 90\%, 70\%, or 50\%.
The asymmetric data-generating mechanism uses smaller pre-peak curvature and larger post-peak curvature, corresponding to slower pre-peak improvement and faster post-peak decline. This scenario is intended as a stress test of trajectory misspecification rather than as a direct reproduction of the empirical pattern observed in the athletics application.

We compare the proposed joint model with two baseline approaches reflecting common practice. The first is a pseudo-peak method (Naive 1), which defines peak age as the age at best observed performance and uses it as the response in a lognormal AFT regression. The second is a two-stage approach (Naive 2). In stage 1, we fit the trajectory model to the longitudinal observations and extract the estimated peak age \(\hat{\tau}_i\) for each athlete. In stage 2, these estimated peak ages are used as responses in the same lognormal AFT regression as Naive 1. Thus, Naive 2 uses the same type of trajectory information as the proposed method, but it first estimates peak ages from fitted trajectories and then treats those estimates as observed outcomes in a separate regression. In contrast, the proposed joint model estimates trajectories, latent peak ages, and AFT regression parameters simultaneously, so uncertainty in \(\tau_i\) is propagated into regression inference.
We evaluate bias, Monte Carlo standard deviation, and average reported standard errors or posterior standard deviations for the regression coefficients. We also assess the accuracy of individual peak-age estimation using interval coverage for $\tau_i$, where applicable. Although inference is Bayesian, these metrics are computed from a
repeated-sampling perspective to examine frequentist operating properties of
the proposed method.

\subsection{Simulation results}

Before presenting the numerical results, we clarify the trajectory-specification labels used in Tables~\ref{tab:sim-all} and~\ref{tab:tau-estimation}. For methods that explicitly model performance trajectories, namely the proposed method and Naive 2, the rows labeled ``Correct'' correspond to data generated from the symmetric quadratic trajectory and fitted with the symmetric trajectory model. The rows labeled ``Wrong'' correspond to data generated from the asymmetric quadratic trajectory but fitted with the symmetric trajectory model. Thus, ``Wrong'' refers specifically to fitting a symmetric trajectory model to asymmetric trajectory data; other conceptually correct combinations, such as fitting the asymmetric model to symmetric or asymmetric data, are not reported.

Table~\ref{tab:sim-all} summarizes the simulation results across the three
settings. For each parameter, entries are reported as the Monte Carlo mean
estimate across replications, with the Monte Carlo standard deviation in
parentheses and the average posterior standard deviation in brackets. Results are shown for both correctly specified and
misspecified trajectory models under varying levels of observational
completeness.

\begin{table}[!ht]
\centering
\scriptsize
\setlength{\tabcolsep}{4pt}
\renewcommand{\arraystretch}{1.15}
\caption{Simulation summaries over 200 replications: Mean (SD) [Mean SE]. Mean and SD are the Monte Carlo mean and standard deviation; Mean SE is the average reported standard error for Naive 1 and Naive 2, and the posterior standard deviation for the proposed method. For the naive AFT regressions, the model-based standard error of $\log\sigma_\tau$ is approximately $1/\sqrt{2I}=0.05$ because $I=200$. Spec denotes the misspecification contrast in Section~\ref{sec:simulation}: Correct is symmetric data fitted with the symmetric trajectory model, and Wrong is asymmetric data fitted with the symmetric trajectory model. The dot for Naive 1 indicates that no trajectory model is fitted. True values are
$\beta_0 = 3.20$, $\beta_1 = 0.01$, $\beta_2 = 0.05$, and $\log\sigma_\tau = -2.30$.} 
\label{tab:sim-all}

\resizebox{\textwidth}{!}{%
\begin{tabular}{llcccc}
\toprule
Method & Spec & $\beta_0$ & $\beta_1$ & $\beta_2$ & $\log\sigma_\tau$ \\
\midrule

\multicolumn{6}{l}{\textit{Setting 1: 90\% peak coverage}} \\

Naive 1 & $\cdot$ 
& 3.206 (0.0280) [0.0280]
& 0.00714 (0.00387) [0.00412]
& 0.03370 (0.03205) [0.03100]
& -1.755 (0.0481) [0.0500] \\

Naive 2 & Correct
& 3.221 (0.0154) [0.0123]
& 0.00557 (0.00200) [0.00183]
& 0.02808 (0.01440) [0.01375]
& -2.572 (0.1065) [0.0500] \\

Naive 2 & Wrong
& 3.160 (0.0133) [0.0099]
& 0.00417 (0.00162) [0.00147]
& 0.02107 (0.01178) [0.01109]
& -2.791 (0.1300) [0.0500] \\

Proposed & Correct
& 3.200 (0.0231) [0.0225]
& 0.00964 (0.00327) [0.00325]
& 0.04997 (0.02539) [0.02487]
& -2.307 (0.0866) [0.0879] \\

Proposed & Wrong
& 3.141 (0.0212) [0.0211]
& 0.00832 (0.00304) [0.00301]
& 0.04295 (0.02337) [0.02321]
& -2.455 (0.0986) [0.0968] \\

\midrule
\multicolumn{6}{l}{\textit{Setting 2: 70\% peak coverage}} \\

Naive 1 & $\cdot$
& 3.167 (0.0257) [0.0259]
& 0.00400 (0.00380) [0.00384]
& 0.02063 (0.02812) [0.02895]
& -1.823 (0.0473) [0.0500] \\

Naive 2 & Correct
& 3.229 (0.0163) [0.0102]
& 0.00369 (0.00179) [0.00151]
& 0.01891 (0.01241) [0.01136]
& -2.772 (0.1703) [0.0500] \\

Naive 2 & Wrong
& 3.160 (0.0149) [0.0083]
& 0.00417 (0.00147) [0.00123]
& 0.01415 (0.01003) [0.00922]
& -2.986 (0.2005) [0.0500] \\

Proposed & Correct
& 3.199 (0.0290) [0.0282]
& 0.00973 (0.00439) [0.00411]
& 0.05139 (0.03180) [0.03126]
& -2.307 (0.1265) [0.1247] \\

Proposed & Wrong
& 3.138 (0.0276) [0.0283]
& 0.00832 (0.00428) [0.00406]
& 0.04295 (0.03027) [0.03102]
& -2.455 (0.1491) [0.1409] \\

\midrule
\multicolumn{6}{l}{\textit{Setting 3: 50\% peak coverage}} \\

Naive 1 & $\cdot$
& 3.120 (0.0245) [0.0238]
& 0.00249 (0.00353) [0.00353]
& 0.01390 (0.02671) [0.02661]
& -1.908 (0.0474) [0.0500] \\

Naive 2 & Correct
& 3.238 (0.0189) [0.0086]
& 0.00266 (0.00163) [0.00128]
& 0.01351 (0.01084) [0.00959]
& -2.963 (0.2810) [0.0500] \\

Naive 2 & Wrong
& 3.169 (0.0187) [0.0075]
& 0.00204 (0.00143) [0.00112]
& 0.01036 (0.00932) [0.00840]
& -3.117 (0.3710) [0.0500] \\

Proposed & Correct
& 3.202 (0.0352) [0.0349]
& 0.01016 (0.00564) [0.00512]
& 0.05325 (0.03860) [0.03863]
& -2.320 (0.2026) [0.1877] \\

Proposed & Wrong
& 3.130 (0.0388) [0.0400]
& 0.00976 (0.00621) [0.00579]
& 0.05254 (0.04224) [0.04415]
& -2.360 (0.2671) [0.2362] \\

\bottomrule
\end{tabular}
}
\end{table}

These two naive methods were implemented in \textsf{R} using standard procedures for AFT regression and mixed-effects modeling. Naive 1 was implemented using \texttt{survival::}\allowbreak\texttt{survreg()} with a lognormal AFT specification, whereas Naive 2 first fitted the mixed-effects quadratic trajectory model in \texttt{Stan} via \texttt{cmdstanr} and then applied the same \texttt{survival::}\allowbreak\texttt{survreg()} lognormal AFT regression to the estimated peak ages.

Three main findings emerge. First, the proposed joint model gives less biased AFT regression estimates across observation-window settings, especially for $\beta_1$ and $\beta_2$. Naive 1 can perform reasonably when the observed window often contains the true peak, but its pseudo-peak response becomes increasingly boundary-driven as peak coverage decreases. Naive 2 uses fitted trajectories and is less sensitive to this boundary effect, but it still treats first-stage peak-age estimates as observed responses.

Second, incomplete observation is reflected differently across methods. In the naive analyses, weaker peak-age information mainly appears as increased bias, while reported standard errors change little or even decrease. In the joint model, the same loss of information appears mainly as larger posterior standard deviations, because uncertainty about latent peak ages is carried into the AFT regression.

Third, trajectory misspecification remains important. All methods deteriorate when asymmetric trajectories are fitted with a symmetric model, but the proposed method retains smaller AFT coefficient bias. As shown below, interval coverage for individual $\tau_i$ is still limited by trajectory-model adequacy.

We also examined the accuracy of individual peak age
estimation. Table~\ref{tab:tau-estimation} summarizes these results over all athletes within each observation-window setting. Across settings, the pseudo-peak approach (Naive 1) has substantially larger RMSE than the trajectory-based methods, reflecting the instability of using the observed best-performance age as a proxy for $\tau_i$. Naive 2 and the proposed method give similar individual peak-age estimates because both use the longitudinal performance records to reconstruct athlete-specific trajectories; nevertheless, the proposed method has slightly smaller RMSE across all settings. 
Under correct trajectory specification, empirical coverage of the 95\% intervals remains close to the nominal level, whereas under trajectory misspecification coverage deteriorates for both trajectory-based methods. The higher coverage in the 50\% misspecified setting likely occurs because weaker peak-age information leads to wider intervals.
Thus, individual $\tau_i$ estimation is ultimately limited by the amount of trajectory information available for each athlete and by trajectory-model adequacy. The advantage of the proposed joint model compared with Naive 2 is therefore not mainly a large improvement in individual peak-age point estimation, but that it propagates uncertainty in the latent peak ages into the AFT regression, whereas Naive 2 treats first-stage peak-age estimates as observed responses. 

\begin{table}[!ht]
\centering
\scriptsize
\setlength{\tabcolsep}{4pt}
\renewcommand{\arraystretch}{1.08}
\caption{Individual latent peak-age estimation across observation-window settings, summarized over all athletes. RMSE is the root mean squared error of posterior mean or point estimates of $\tau_i$. Cov. is the empirical coverage of 95\% intervals for $\tau_i$ over 200 replications. Spec uses the same misspecification contrast as Table~\ref{tab:sim-all}. Coverage is not applicable for Naive 1 because it does not produce individual peak-age intervals.}
\label{tab:tau-estimation}
\begin{tabular}{llcccccc}
\toprule
& & \multicolumn{2}{c}{Naive 1} & \multicolumn{2}{c}{Naive 2} & \multicolumn{2}{c}{Proposed} \\
\cmidrule(lr){3-4}\cmidrule(lr){5-6}\cmidrule(lr){7-8}
Window & Spec & RMSE & Cov. & RMSE & Cov. & RMSE & Cov. \\
\midrule
\multicolumn{8}{l}{\textit{Correct trajectory specification}} \\
90\% & Correct & 4.195 & -- & 1.769 & 0.947 & 1.721 & 0.948 \\
70\% & Correct & 4.410 & -- & 2.164 & 0.946 & 2.096 & 0.946 \\
50\% & Correct & 4.780 & -- & 2.393 & 0.944 & 2.333 & 0.945 \\
\midrule
\multicolumn{8}{l}{\textit{Misspecified trajectory specification}} \\
90\% & Wrong & 4.311 & -- & 2.565 & 0.800 & 2.516 & 0.795 \\
70\% & Wrong & 4.541 & -- & 2.810 & 0.805 & 2.744 & 0.804 \\
50\% & Wrong & 4.982 & -- & 2.994 & 0.866 & 2.933 & 0.867 \\
\bottomrule
\end{tabular}
\end{table}

\section{Discussion}
\label{sec:discussion}

This work develops a hierarchical Bayesian joint model for inferring latent peak ages and performance trajectories from sparse, noisy, and irregular competition records. By treating peak age as a latent event time defined by the underlying trajectory, the framework links performance dynamics directly to covariate effects on earlier or later peak age.

Methodologically, the main contribution is to connect longitudinal trajectory modeling with event-time inference when the event, peak performance, is not observed directly but defined as a functional of the latent trajectory. This formulation propagates trajectory uncertainty into peak-age and covariate-effect inference, addressing a key limitation of pseudo-peak and two-stage analyses. In simulations, as peak-age information weakens, pseudo-peak and two-stage analyses tend to attenuate regression effects while reporting standard errors that change little or even decrease. The joint model instead reflects weaker information through wider posterior uncertainty and comparatively stable regression estimates. This pattern echoes two familiar statistical issues: measurement error in a latent response can bias covariate effects, and two-stage generated-response procedures can make second-stage inference overconfident when first-stage uncertainty is ignored.

In the athletics application, the framework gives coherent inference on peak age and on rates of improvement and decline while accounting for heterogeneity across athletes and disciplines. The results show systematic sport-category differences in peak age and trajectory shape, along with biological-sex patterns that are not fully explained by conventional physiological arguments. These findings show how statistical modeling can complement domain knowledge about biological development, career structure, and selection processes.

Beyond this application, the framework has broader use in sports analytics and related fields. It can support athlete development analysis, forecasting from partial career data, and comparisons across sports or cohorts where peak age and performance dynamics may reflect structural or environmental differences. More generally, the methodology applies when a latent event time is directly defined by an underlying longitudinal process, such as growth curves, skill acquisition, or productivity trajectories.

Several extensions and data-related limitations suggest future work. The parametric AFT submodel is one convenient specification for latent peak age; semi-parametric or nonparametric alternatives could be developed within the same joint framework, although integration with the latent trajectory model remains nontrivial. Similarly, quadratic trajectories are interpretable but may not capture multi-phase or multi-peak career patterns, motivating smoother and more flexible trajectory representations. Data-related limitations include the lack of explicit information on injury, retirement, event transitions, competition-level changes, and competing career endpoints, all of which may inform trajectories and peak age. Because the World Athletics analysis was restricted to athletes in a single selected event, future work could model cross-event dependence and event-transition processes for multi-discipline athletes. Extensions to multivariate performance outcomes would further broaden the framework.

For reproducibility, we provide an accompanying \textsf{R} package supporting symmetric and asymmetric trajectory specifications. The software is available via an anonymized GitHub repository and will be publicly released.

\section*{Funding}
T. Guan's work was supported by the Discovery Grant (RGPIN-2022-05140) from the Natural Sciences and Engineering Research Council of Canada (NSERC).

\section*{Supplementary Material}
Additional figures and tables, simulation details, prior specifications, and MCMC diagnostics are provided in the accompanying supplementary material.

\bibliographystyle{plainnat}
\begingroup
\small
\bibliography{Bibliography-MM-MC}
\endgroup

\clearpage


\section*{Supplementary material}

# Supplementary Material: Hierarchical Bayesian Joint Modeling of Athlete Performance Trajectories and Latent Peak Ages

K. Ken Peng[a,b], Tianyu Guan[c], and Xiong Zhao[d,e]

[a]Department of Statistics and Actuarial Science, Simon Fraser University, Burnaby, BC, Canada
[b]Department of Civil Engineering, University of Ottawa, Ottawa, ON, Canada
[c]Department of Mathematics and Statistics, York University, Toronto, ON, Canada
[d]AccMov Health Inc., Ottawa, ON, Canada
[e]School of Kinesiology and Health Studies, Queen's University, Kingston, ON, Canada

## S1 Additional figures and tables

Table S1 provides the event-level summary of the World Athletics dataset after preprocessing.

Table S1: World Athletics dataset after preprocessing: 3,229 single-event athletes and 105,515 competition records. Observed window is the last observed age minus the first observed age for each athlete. IQR denotes the interquartile range. Observed age range reports the median first and last observed ages within each event.

| Event | No. athletes | No. records | Records/athlete median (IQR) | Observed window median (IQR) | Observed age range median |
|---|---|---|---|---|---|
| 100 m | 470 | 15,000 | 22 (14.0, 41.0) | 6.9 (4.2, 10.0) | 19.5–27.7 |
| 400 m | 606 | 18,537 | 22 (13.0, 39.0) | 5.3 (3.3, 8.3) | 20.9–27.0 |
| 1500 m | 639 | 16,647 | 20 (13.0, 33.0) | 6.4 (4.2, 9.0) | 22.4–29.3 |
| Marathon | 238 | 3,039 | 12 (11.0, 14.0) | 9.0 (7.0, 10.5) | 24.7–34.0 |
| Long jump | 286 | 8,491 | 22 (14.0, 36.0) | 7.4 (5.2, 10.1) | 21.0–29.2 |
| Shot put | 379 | 17,932 | 30 (17.0, 59.5) | 6.8 (4.4, 9.4) | 22.8–30.1 |
| Discus throw | 213 | 9,189 | 26 (15.0, 49.0) | 8.3 (6.0, 12.0) | 23.0–32.0 |
| Javelin throw | 366 | 16,204 | 29 (17.0, 57.0) | 7.7 (5.0, 10.7) | 21.4–29.8 |
| Decathlon | 13 | 200 | 13 (11.0, 19.0) | 7.2 (6.2, 8.2) | 21.0–29.1 |
| Heptathlon | 19 | 276 | 14 (12.0, 17.0) | 8.0 (7.0, 9.0) | 22.5–30.6 |
| Total | 3,229 | 105,515 | – | – | – |

Figure S1 summarizes the athlete-level posterior mean peak ages on the original age scale by sport category and biological sex.

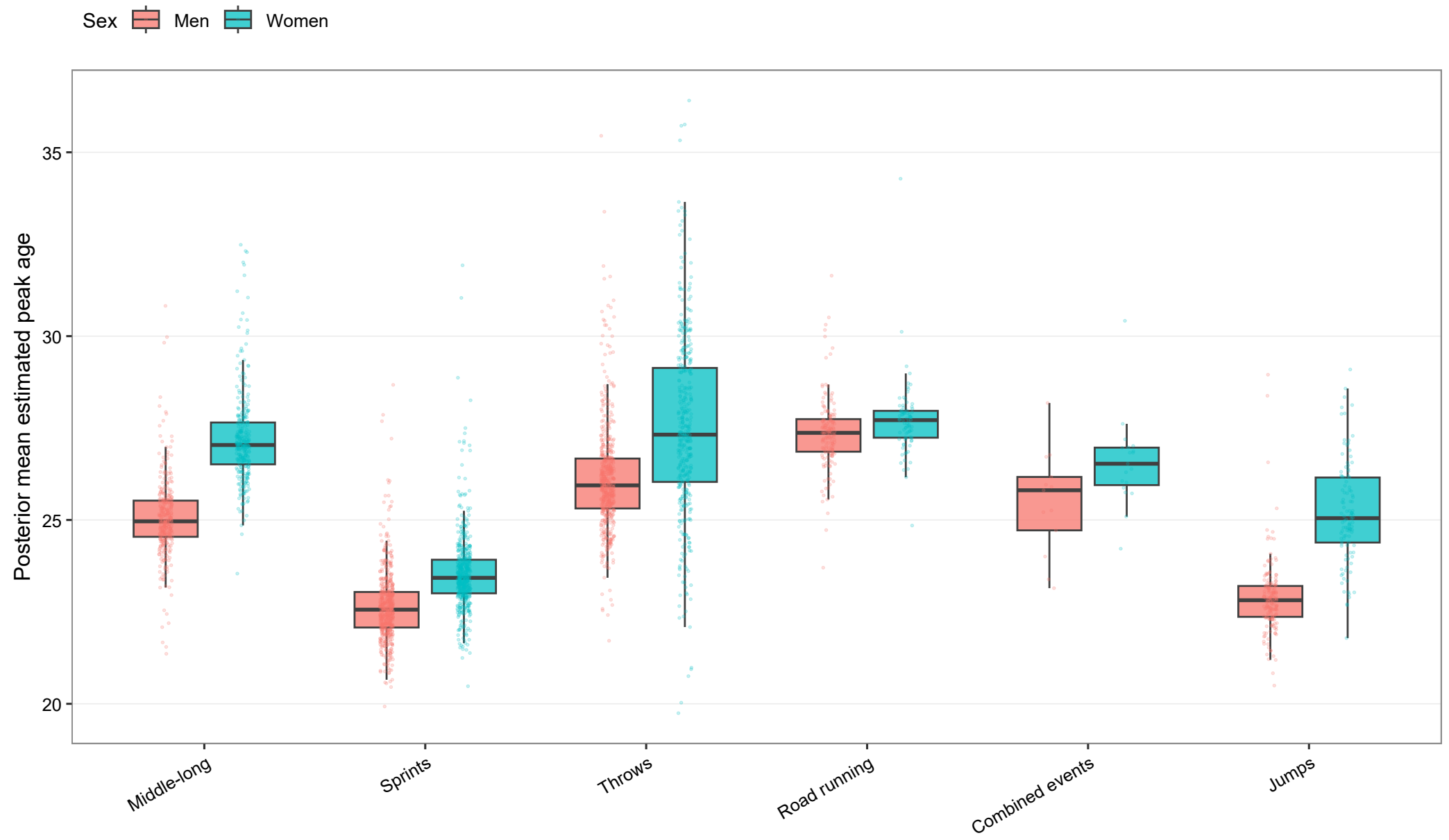


Figure S1: Athlete-level posterior mean peak ages by sport category and biological sex. Boxplots summarize posterior mean estimates of latent peak age $\tau_i$ within each sport–sex group; points represent individual athletes.

# S2 Simulation data-generating procedure

Algorithm S1 gives the full data-generating procedure used in the simulation study.

**Algorithm S1** Simulation data-generating procedure.

Fix $I = 200$ athletes and repeat for $M = 200$ replications. Fix $\beta_0 = 3.20$, $\beta_1 = 0.01$, $\beta_2 = 0.05$, $\sigma_\tau = 0.10$, $\alpha_0 = 1150$, $\sigma_\alpha = 50$, $\rho_0 = 0.25$, $\sigma_\rho = 0.8$, and $\sigma_y = 20$.

**for** $i = 1, \dots, I$ **do**

Generate covariates $x_{1i} \sim N(0, 3^2)$, $x_{2i} \sim \text{Bernoulli}(0.8)$.

Generate peak age:

$$\log(\tau_i) = \beta_0 + \beta_1 x_{1i} + \beta_2 x_{2i} + e_i, \quad e_i \sim N(0, \sigma_\tau^2).$$

Generate trajectory parameters:

$$\alpha_i \sim N(\alpha_0, \sigma_\alpha^2), \qquad \rho_i \sim \text{Lognormal}(\log \rho_0, \sigma_\rho^2).$$

Construct the symmetric quadratic trajectory:

$$Y_i(t) = \alpha_i - \rho_i (t - \tau_i)^2.$$

For the asymmetric data-generating mechanism, use

$$Y_i(t) = \begin{cases} \alpha_i - 0.5\rho_i (t - \tau_i)^2, & t \le \tau_i, \\ \alpha_i - 1.5\rho_i (t - \tau_i)^2, & t > \tau_i. \end{cases}$$

Generate independent $U_i, V_i \sim \text{Beta}(2, 5)$. Let $L_i$ denote the start of the observed age window, $S_i$ its length, and $R_i$ its endpoint. Set $(L_i, S_i) = (15 + 10U_i, 8 + 40V_i)$, $(15 + 15U_i, 40V_i)$, or $(15 + 15U_i, 25V_i)$ for the approximately 90%, 70%, and 50% peak-coverage settings, respectively, and then set $R_i = \min(45, L_i + S_i)$.

Draw $n_i = \max\{3, \text{Poisson}(20)\}$, set two observation times to $L_i$ and $R_i$, and sample the remaining $n_i - 2$ times independently from $\text{Uniform}(L_i, R_i)$.

Generate observations:

$$y_{ij} = Y_i(t_{ij}) + \varepsilon_{ij}, \quad \varepsilon_{ij} \sim N(0, \sigma_y^2).$$

**end for**

# S3 Prior specification

Table S2 summarizes the prior distributions used for the athletics application. These priors are weakly informative: they encode broad plausible ranges for elite athletics performance while leaving substantial room for posterior updating by the data.

Table S2: Prior distributions for the athletics application.

| Parameter block | Prior distribution | Hyperparameters |
|---|---|---|
| Components of $\boldsymbol{\theta}_0$ corresponding to peak level | Normal | mean = 1200, sd = 200 |
| Components of $\boldsymbol{\theta}_0$ corresponding to trajectory curvature[a] | Normal | mean = log(0.5), sd = 1 |
| Sport-category coefficients in $\mathbf{B}$ for peak level | Normal | mean = 0, sd = 200 |
| Sport-category coefficients in $\mathbf{B}$ for trajectory curvature | Normal | mean = 0, sd = 0.5 |
| $\beta_0$ | Normal | mean = log(28), sd = 0.5 |
| Biological-sex coefficient in $\boldsymbol{\beta}$ | Normal | mean = −0.1, sd = 0.1 |
| Birth-year coefficient in $\boldsymbol{\beta}$ | Normal | mean = 0, sd = 0.5 |
| Sport-category coefficients in $\boldsymbol{\beta}$ for peak age | Normal | domain-informed means[b], sd = 0.1 |
| Sport-by-sex interaction coefficients in $\boldsymbol{\beta}$ | Normal | mean = 0, sd = 0.3 |
| Variance components in $\boldsymbol{\Sigma}_\theta$[c] | Half-Normal | sd = 100 |
| $\sigma_y$ | Half-Normal | sd = 200 |
| $\sigma_\tau$ | Half-Normal | sd = 0.5 |

[a] In the implementation, curvature parameters are modeled on the log scale to ensure positivity.

[b] Domain-informed prior means for sport-category effects in the peak-age model are 0.10 (combined-events), −0.10 (sprints), 0.00 (throws), −0.10 (road running), and 0.00 (jumps), all specified relative to the baseline middle-long distance category.

[c] In the present implementation, $\boldsymbol{\Sigma}_\theta$ is assumed diagonal and parameterized through independent random-effect scale parameters for computational efficiency.

# S4 MCMC diagnostics

Posterior inference for the athletics application was carried out using four parallel MCMC chains. Figures S2–S5 show trace plots and prior-to-posterior comparisons for representative parameters. In the trace plots, the four chains overlap closely and fluctuate around stable levels without visible long-term trends, indicating good mixing and stationarity for both the peak-age AFT submodel and the longitudinal performance submodel. The reported $\widehat{R}$ values are close to one for all displayed parameters. The $\widehat{R}$ diagnostic compares between-chain and within-chain variability, with values near one indicating that the chains are sampling from the same stationary posterior distribution (Vehtari et al., 2021). In these diagnostics, the displayed $\widehat{R}$ values range from 1.001 to 1.007, providing no evidence of lack of convergence.

The prior-to-posterior comparisons show that the data provide substantial information beyond the weakly informative priors. For the peak-age AFT parameters, the posterior densities are generally much more concentrated than the corresponding priors. Several posterior distributions also shift away from their prior centers, for example for the biological-sex effect, the road-running sport-category effect, and selected sport-by-sex interaction terms. For the trajectory-curvature parameters, the posterior densities are likewise narrower than the priors and are centered differently across sport categories and between the pre-peak and post-peak curvature components. These patterns indicate that the posterior summaries are driven by the observed longitudinal performance records rather than being dominated by the prior specification.

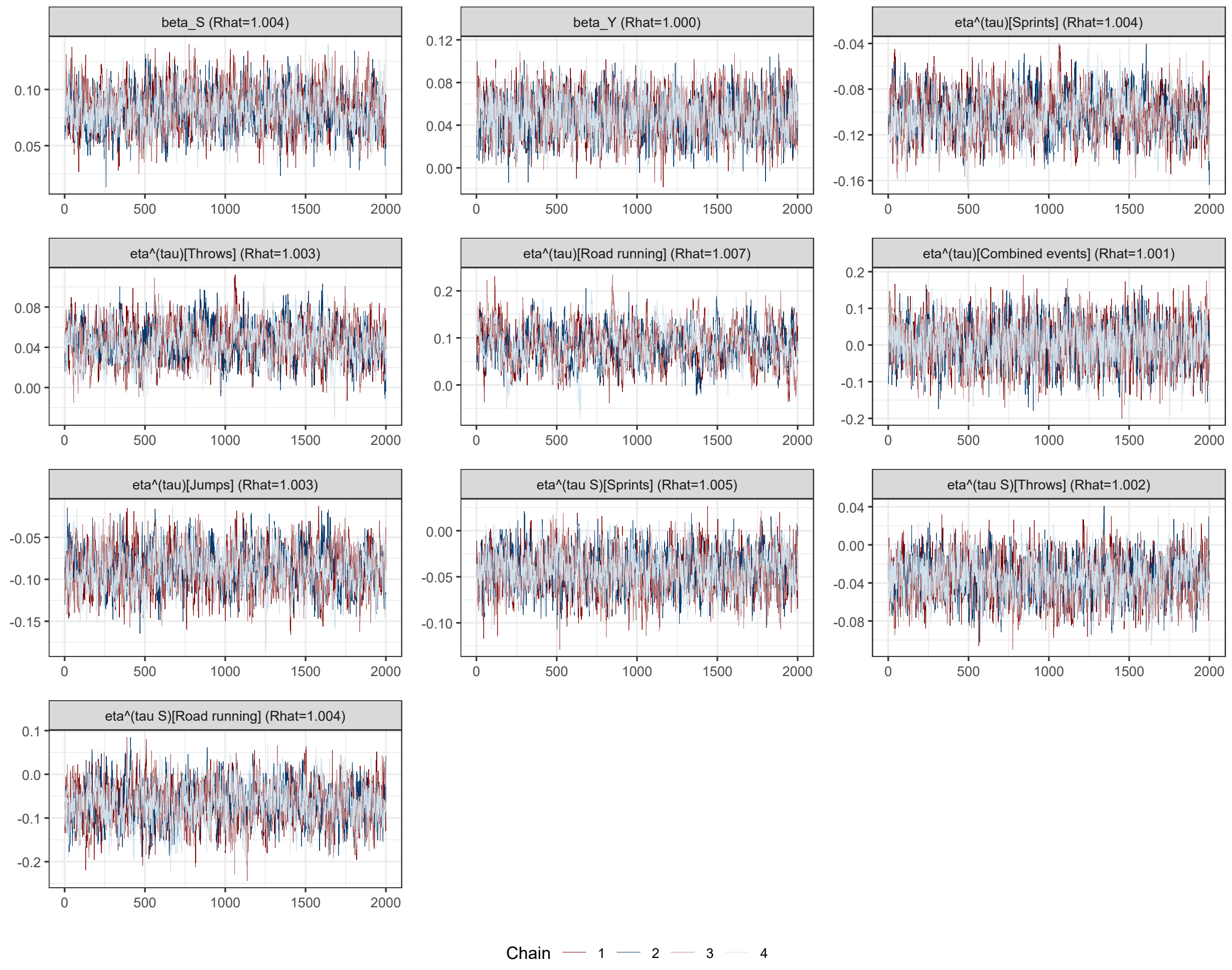


Figure S2: Trace plots for selected peak-age AFT parameters.

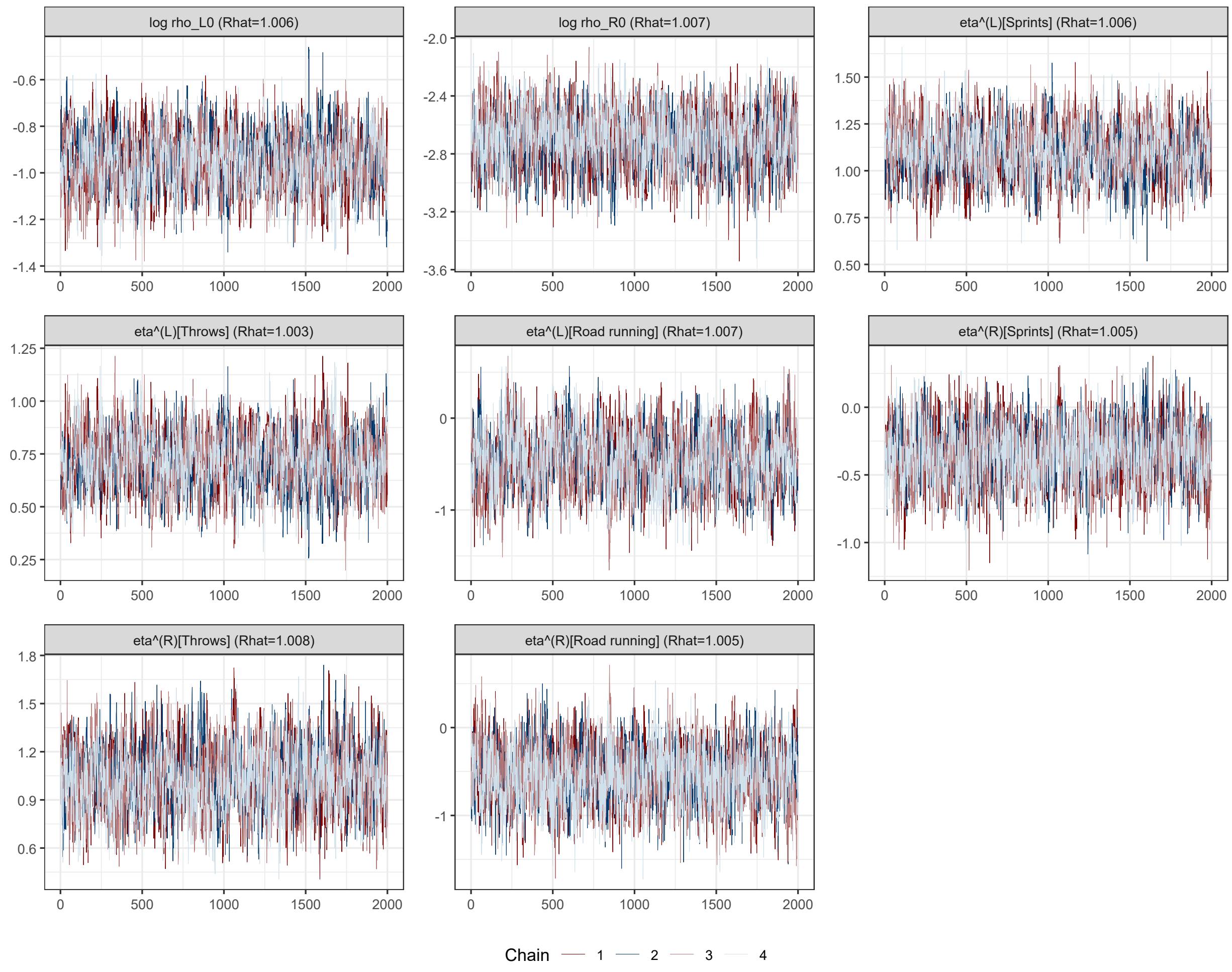


Figure S3: Trace plots for selected longitudinal performance parameters.

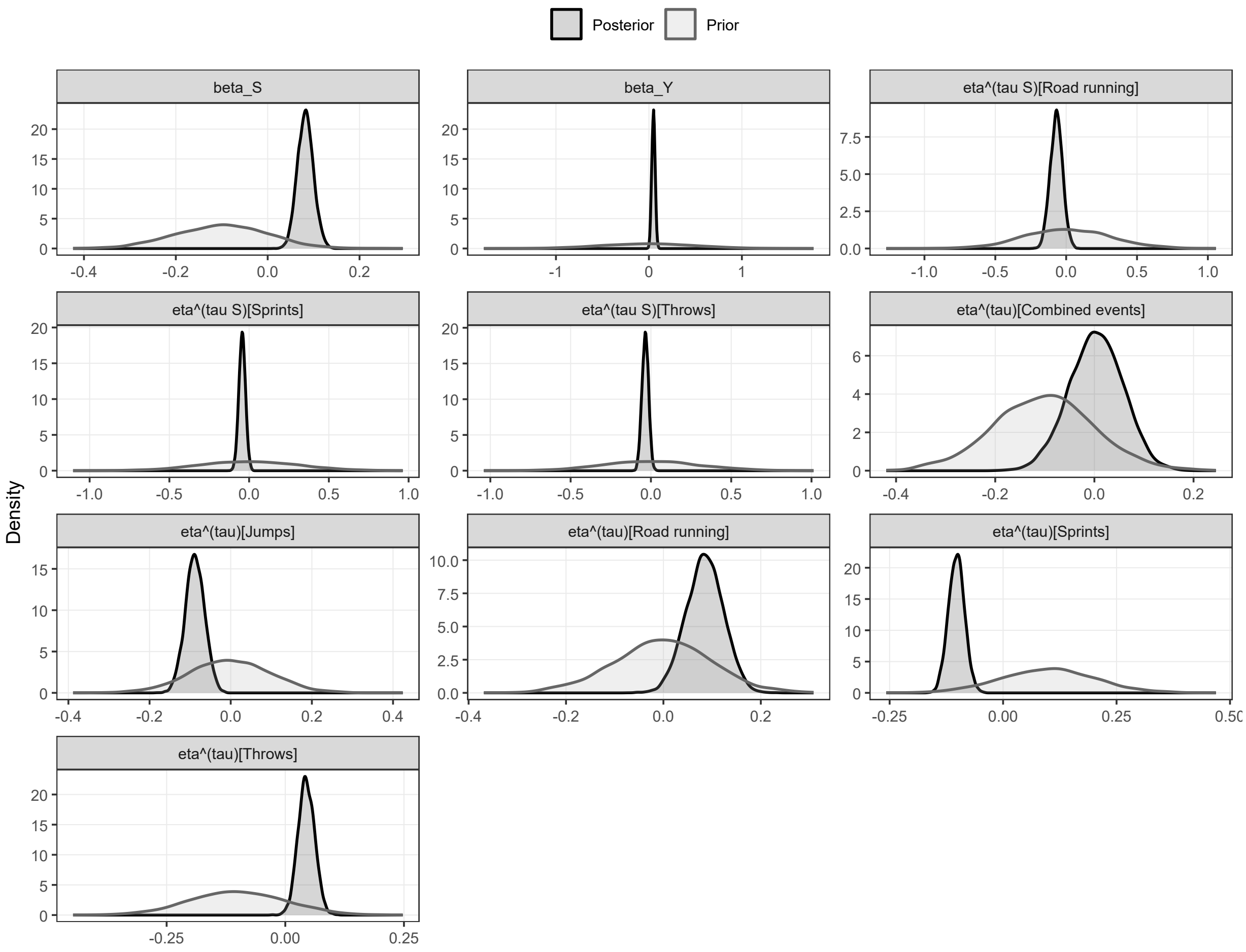


Figure S4: Prior and posterior densities for selected peak-age AFT parameters.

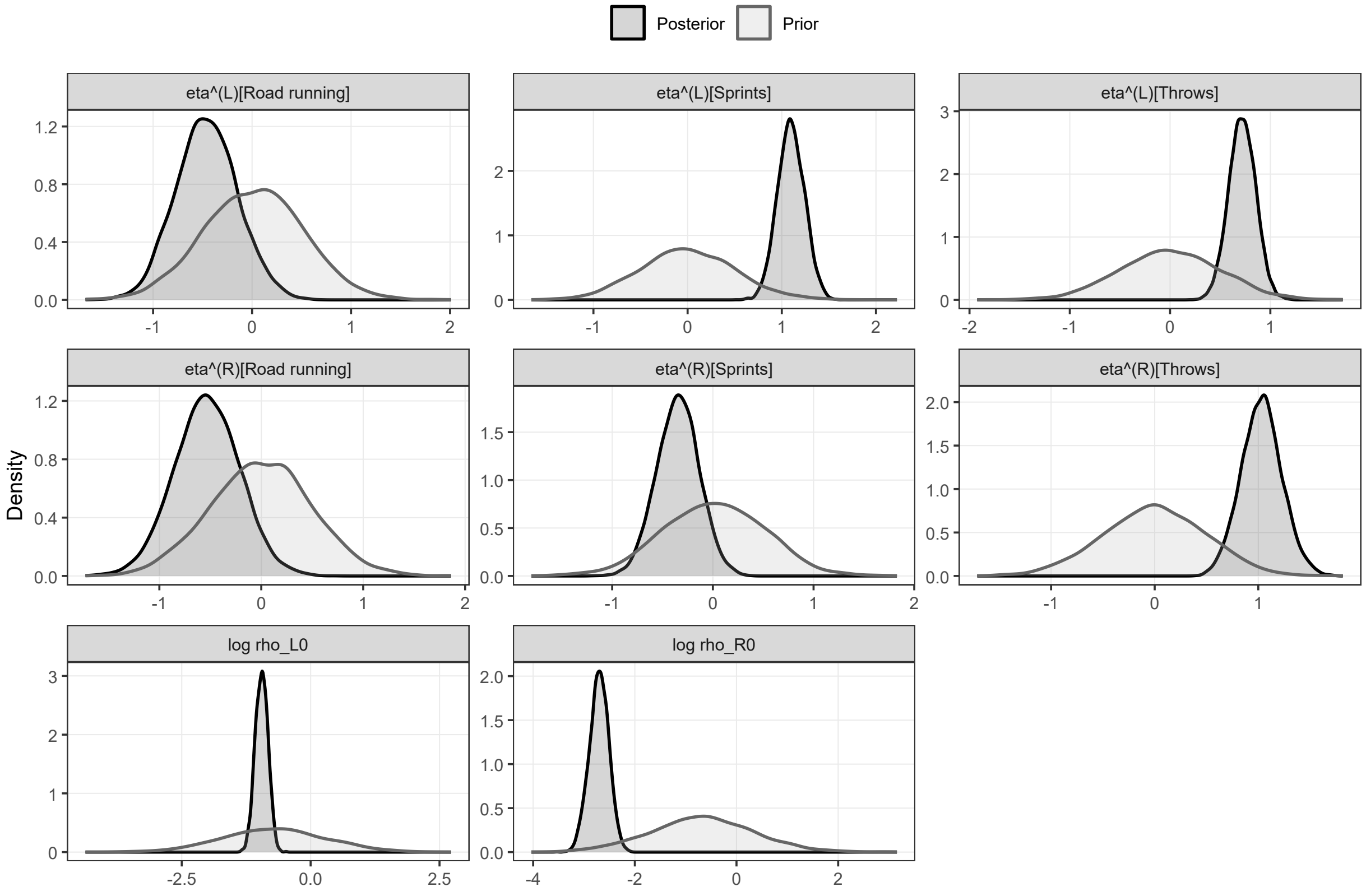


Figure S5: Prior and posterior densities for selected trajectory-curvature parameters.



\end{document}